\documentclass[%
reprint,
superscriptaddress,
nofootinbib,
 amsmath,amssymb,
prd,
floatfix,
]{revtex4-1}
\usepackage[dvipsnames]{xcolor}

\usepackage[utf8]{inputenc}
\usepackage{graphicx}
\usepackage{xcolor}
\usepackage{comment}
\usepackage{soul}
\usepackage{siunitx}
\usepackage{xspace}
\usepackage[mathlines]{lineno}

\newcommand{\eeff}{\epsilon_{\mathrm{eff}}}
\newcommand{\eh}{e$^-$/h$^+$\xspace}

\begin{document}

\title{Investigating the sources of low-energy events in a SuperCDMS-HVeV detector}

\author{M.F.~Albakry} \affiliation{Department of Physics \& Astronomy, University of British Columbia, Vancouver, BC V6T 1Z1, Canada}\affiliation{TRIUMF, Vancouver, BC V6T 2A3, Canada}
\author{I.~Alkhatib} \affiliation{Department of Physics, University of Toronto, Toronto, ON M5S 1A7, Canada}
\author{D.W.P.~Amaral} \affiliation{Department of Physics, Durham University, Durham DH1 3LE, UK}
\author{T.~Aralis} \affiliation{Division of Physics, Mathematics, \& Astronomy, California Institute of Technology, Pasadena, CA 91125, USA}
\author{T.~Aramaki} \affiliation{Department of Physics, Northeastern University, 360 Huntington Avenue, Boston, MA 02115, USA}
\author{I.J.~Arnquist} \affiliation{Pacific Northwest National Laboratory, Richland, WA 99352, USA}
\author{I.~Ataee~Langroudy} \affiliation{Department of Physics and Astronomy, and the Mitchell Institute for Fundamental Physics and Astronomy, Texas A\&M University, College Station, TX 77843, USA}
\author{E.~Azadbakht} \affiliation{Department of Physics and Astronomy, and the Mitchell Institute for Fundamental Physics and Astronomy, Texas A\&M University, College Station, TX 77843, USA}
\author{S.~Banik} \affiliation{School of Physical Sciences, National Institute of Science Education and Research, HBNI, Jatni - 752050, India}
\author{C.~Bathurst} \affiliation{Department of Physics, University of Florida, Gainesville, FL 32611, USA}
% ...

\date{\today}

\begin{abstract}
Recent experiments searching for sub-GeV/$c^2$ dark matter have observed event excesses close to their respective energy thresholds. Although specific to the individual technologies, the measured excess event rates have been consistently reported at or below event energies of a few-hundred eV, or with charges of a few electron-hole pairs. In the present work, we operated a 1-gram silicon SuperCDMS-HVeV detector at three voltages across the crystal (0\,V, 60\,V and 100\,V). The 0\,V data show an excess of events in the tens of eV region. Despite this event excess, we demonstrate the ability to set a competitive exclusion limit on the spin-independent dark matter--nucleon elastic scattering cross section for dark matter masses of $\mathcal{O}(100)$ MeV/$c^2$, enabled by operation of the detector at 0\,V potential and achievement of a very low $\mathcal{O}(10)$\,eV threshold for nuclear recoils. Comparing the data acquired at 0\,V, 60\,V and 100\,V potentials across the crystal, we investigated possible sources of the unexpected events observed at low energy. The data indicate that the dominant contribution to the excess is consistent with a hypothesized luminescence from the printed circuit boards used in the detector holder.

\end{abstract}

\maketitle


\section{Introduction}

Low-mass (sub-GeV/$c^2$) dark matter searches have benefited from detector development with low energy threshold and low readout noise. Despite this progress, their reach has been challenged by unexpected, excess event rates. These include reports from experiments using cryogenic calorimeters instrumented for readout of phonon signals, such as EDELWEISS~\cite{Armengaud:2019,Arnaud:2020}, CRESST~\cite{Abdelhameed:2019}, NuCLEUS~\cite{Angloher:2017, rothe:2019}, and SuperCDMS-CPD~\cite{Alkhatib:2020cpd}. Unexpected events are also present in detectors instrumented for charge readout, such as the CCD-based experiments SENSEI~\cite{Barak:2020} and DAMIC~\cite{Aguilar:2020}, as well as the phonon-based measurement of ionization signals~\cite{Agnese:2018, Amaral:2020ryn}.

These latter measurements were made possible by the development by the SuperCDMS Collaboration of silicon-based gram-scale detectors: the high-voltage eV-resolution (HVeV) detectors~\cite{Romani:2018,ren:2020}. These detectors can be operated in high voltage (HV) mode in which an applied electric field amplifies the signal from electron-hole pairs (\eh) via the Neganov-Trofimov-Luke (NTL) effect~\cite{Neganov:1985,Luke:1988}. If the voltage is sufficiently high, the signal represents the number of \eh, and a trigger threshold of well below a single \eh was reached. However, these devices can also be operated in zero-voltage (0V) mode. In this case the measured signal represents the actual interaction energy.

We undertook an above-ground search for dark matter with a second-generation Si HVeV detector. 
An analysis of the data taken in the HV mode (100\,V) was described in Ref.~\cite{Amaral:2020ryn} and measured an unexplained excess of events similar to those observed with a previous version of the detector~\cite{Agnese:2018}. In order to better understand this excess event rate, in this manuscript we analyze the data taken in the 0V mode alongside the data taken at two different high-voltage settings: 60\,V and 100\,V. We infer information about the origin of the observed events by comparing how the spectrum scales with the applied voltages.

This manuscript is arranged as following: We review the experimental setup in Sec.~\ref{sec:2-setup} and present the event reconstruction algorithms in Sec.~\ref{sec:3-reconstruction}. We present a dark matter analysis of the data taken in the 0V mode in Sec.~\ref{sec:4-DM}. The investigation of the low-energy events starts in Sec.~\ref{sec:5-pulse_shape}, where we discuss a class of events with anomalous pulse shape found in the dark matter search data, and in Sec.~\ref{sec:6-compare} we compare the pulse shapes and energy spectra from data taken at different voltages. In Sec.~\ref{sec:discussion}, we discuss a plausible explanation for the low-energy events with the anomalous pulse shape.

\section{Experimental setup and Data Collection}\label{sec:2-setup}

The experimental setup and data collection conditions used in this analysis are identical to those described in Ref.~\cite{Amaral:2020ryn}. Details pertinent to this report follow. The detector substrate is a 0.93\,gram high-purity Si crystal with dimensions of $1.0\times1.0\times0.4$\,cm$^{3}$. Two distributed channels of Quasiparticle-trap-assisted Electrothermal-feedback TESs \footnote{TESs = Transition Edge Sensors}~\cite{Irwin:2005, ren:2020} (QETs) are patterned on the front surface to measure phonon signals. An aluminum grid is patterned on the back surface to enable application of a voltage bias across the crystal. Two printed circuit boards (PCBs) clamp the detector for thermal sinking and to facilitate electrical connections. The QETs are connected via wirebonds to traces on the PCB top surface. A light-tight copper box surrounds the detector and the PCB clamps. The detector is deployed in an Adiabatic Demagnetization Refrigerator (ADR).
A continuous data acquisition system digitized detector signals at a sampling frequency of 1.51\,MHz.

We collected data during April 29--May 16 of 2019, including calibration data and dark matter (DM) search data at 0\,V, 60\,V and 100\,V. Each day during the data-taking campaign, the ADR was recooled down from above 4~K. The ADR base temperature was stabilized at 50\,mK from April 29th to May 7th and at 52\,mK from May 8th to May 16th. Both channels of QETs were operated at 45\% of their normal-state resistance. We calibrated the detector energy response daily using a 635 nm laser that was fiber-coupled from room temperature to the detector housing. We also took calibration data on May 14 with an $^{55}\textrm{Fe}$ source at crystal biases up to 60\,V to extend the detector calibration to $\sim$100\,keV.

\section{Event Reconstruction}\label{sec:3-reconstruction}

In this section we describe the triggering and energy-reconstruction algorithms, and the energy calibration procedure. In-depth discussions of the algorithms and calibration procedure can be found in Refs.~\cite{Amaral:2020ryn,ren:2020}.

\subsection{Triggering and energy reconstruction}\label{subsec:3a-reconstruction}

We read out a continuously sampled timestream of the current flowing through each QET detector channel. The sum of the traces from the two channels is filtered with an optimum filter (OF)~\cite{Gatti:1986, Kurinsky:2018} before applying the threshold trigger as part of the offline data processing. In this analysis we use an OF time window of 10.8\,ms, with equal pre- and post-trigger regions of 5.4\,ms. We build a pulse template for the OF using events with a total phonon energy of  $\sim$100\,eV from laser-calibration data, an energy deposition where the detector is far from its saturation regime and thus its response is expected to be linear. We also calculate the noise power spectral density (PSD) on a daily basis from randomly-selected sections of the data that lacks pulses. We set a 9.2\,eV trigger threshold for the dark matter constraints discussed in Sec.~\ref{sec:4-DM}, which results in a 20\,Hz trigger rate. For the comparison of 0V and HV data discussed in Sec.~\ref{sec:5-pulse_shape} and onward, we use a higher threshold of 15\,eV to reduce the contribution from triggers caused by noise.

We use the amplitudes calculated by the OF algorithm as the energy estimator for low energy events, and use an integral-based energy estimator for high-energy events. At higher energies the TESs approach their normal-state resistances, resulting in ``flat-topped" pulses. These saturated pulses have shapes that deviate significantly from the pulse template, resulting in degradation of the energy sensitivity of the OF amplitude. The integral-based energy estimator integrates over the raw trace when the detector is saturated and the signal-to-noise ratio is high, and integrates the area below a fit to the tail of the pulse using the average pulse template where the signal-to-noise ratio is low. We refer to this estimator as the ``Matched Filter (MF) integral"~\cite{ren:2020}. The detector energy reconstruction is based on the OF amplitude below 600\,eV and MF integral above 800\,eV, with a linear transition in between.

\subsection{Energy calibration}\label{subsec:3b-calibration}

In this section, we discuss the calibration procedure using HV data, the application of the daily gain corrections, and their combination into a single calibrated energy estimator. We also discuss how this calibration is applied to 0V data in the end.

We calibrate the detector from the threshold to $\sim$\,100\,keV. The calibration is divided into two parts: (1) low-energy calibration using a laser, up to 700\,eV at 100\,V bias; (2) high-energy calibration using a combination of laser data up to 700\,eV and $^{55}\mathrm{Fe}$ source data up to 104\,keV with bias voltages up to 60\,V. We collected laser data every day for robust low-energy calibrations that accounts for the daily gain change due to thermal cycling of the ADR. In contrast, it was not practical to conduct daily high-energy calibration, because of the extended source exposure required to acquire sufficient event statistics. We, therefore, took the high-energy calibration data only once during the data-taking campaign on a dedicated day (``Fe-day").

\begin{figure}[htbp]
    \centering
    \includegraphics[width=0.48\textwidth]{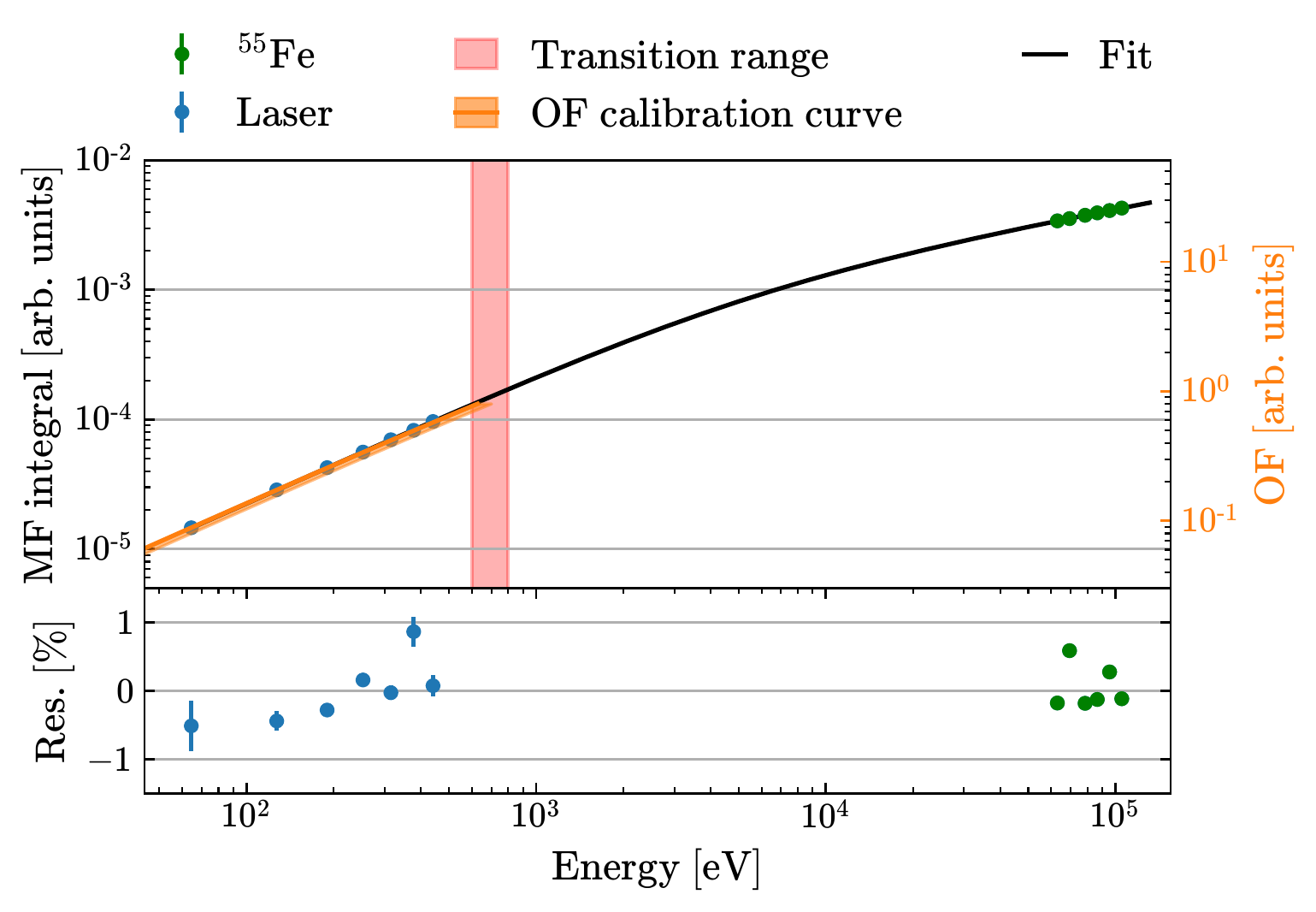}
    \caption{Top: Application of the energy calibration curve. The MF integral calibration curve (black) and a representative example of the OF calibration curve from Fe-day (orange, corresponds to Eq.~\ref{eqn:of}). The OF calibration curve includes an $\sim$11\% systematic uncertainty band, and corresponds to the y-axis on the right side. The vertical red shaded region marks the 600--800\,eV transition range. Bottom: residuals between the data points and the calibration curve expressed as a percentage.}
    \label{fig:calibration}
\end{figure}

The low-energy calibration follows a similar method as described in Ref.~\cite{Amaral:2020ryn}. We use laser data to calculate calibration functions $E_{\mathrm{OF},i}$ to convert OF amplitudes ($A_{\mathrm{OF}}$) to energies up to 700\,eV. The subscript $i$ denotes the $i^{\textrm{th}}$ day of data-taking. The function is a second order polynomial
\begin{equation}
\label{eqn:of}
  E_{\mathrm{OF},i} = a_i\cdot A_{\mathrm{OF}}+b_i\cdot A_{\mathrm{OF}}^2, 
\end{equation}
where $a_i$ and $b_i$ are the two calibration coefficients for the $i^{\mathrm{th}}$ day. An example of the OF calibration curve is shown in Fig.~\ref{fig:calibration}.

We derive a second calibration function based on the MF integral up to 98\,keV with the laser data and the $^{55}\mathrm{Fe}$ data at 40\,V and 50\,V as well as 60\,V with data at the additional voltages used to map out the non-linearity in the high-energy range. The $^{55}\mathrm{Fe}$ source emits X-rays with two characteristic energies of 5.9\,keV and 6.5\,keV~\cite{williams2001x}. The total phonon energy of the two characteristic lines at the applied voltages are calculated according to Ref.~\cite{ren:2020}. We use a $4^{\mathrm{th}}$-order polynomial to model the MF integral as a function of the total phonon energy, as shown in Fig.~\ref{fig:calibration}. This parameterization is used to accommodate the high-energy data points which suffer from saturation effects. These effects cannot be described by a $
2^{\mathrm{nd}}$-order polynomial as they are intrinsically of higher order, driven by the response of a TES to large energy depositions. The resulting calibration function is denoted as  $E_{\mathrm{MF,Fe}}$, where the subscript ``Fe'' specifies that it is derived from data acquired on the dedicated high-energy calibration day.

To account for daily variation of the detector working point relative to Fe-day, we calculate a day-by-day correction factor as the ratio of the low-energy calibration's linear-term coefficient between the $i^{\mathrm{th}}$ and Fe-day: $k_i=\frac{a_i}{a_{\mathrm{Fe}}}$.
We then scale the calibration function based on the MF integral by this multiplicative factor. The corrected high-energy calibration function for the $i^{\mathrm{th}}$ day is $E_{\mathrm{MF},i} = k_i \cdot E_{\mathrm{MF,Fe},}$.

We combine the low-energy and high-energy estimators with the smooth transition shown in Eq.~(\ref{eqn:calibration_combine}),
\begin{equation}
\label{eqn:calibration_combine}
    E_{\mathrm{ph}}= 
\begin{cases}
    E_{\mathrm{OF}},			& E_{\mathrm{OF}}<600\,\mathrm{eV}\\
    (1-c)E_{\mathrm{OF}} + c\cdot E_{\mathrm{MF}}, & 600\,\mathrm{eV} \leq E_{\mathrm{OF}} \leq 800\,\mathrm{eV}\\
	E_{\mathrm{MF}},			& 800\,\mathrm{eV}<E_{\mathrm{OF}}
\end{cases}
\end{equation}
in which $c=\frac{E_{\mathrm{OF}}-600\,\mathrm{eV}}{200\,\mathrm{eV}}$. $E_\mathrm{ph}$ is the calibrated total phonon energy of an event, and $E_{\mathrm{OF}}$ and $E_{\mathrm{MF}}$ are the energy of an event calculated by the low-energy and high-energy calibrations, respectively.

We note that the above calibration is derived with data collected in the HV mode. Ref.~\cite{ren:2020} shows that the calibration of the same detector for the 0V mode can be different from that for the HV mode by $\sim$11\%. For this study, we use the above described calibration for both HV-mode and 0V-mode data, and use the difference as the systematic uncertainty of the calibration (as shown in Fig.~\ref{fig:calibration}). As shown in Sec.~\ref{sec:6-compare}, this systematic uncertainty is negligible in the comparison of the 0V and HV mode data.

\section{Dark matter constraints with 0V spectrum}\label{sec:4-DM}

In this section we consider the energy spectrum with zero bias across the crystal to constrain the the spin-independent DM--nucleon elastic scattering cross section. We describe the live-time and event-based selection criteria and present the dark matter exclusion limit.

\subsection{Live-time estimate}\label{subsec:4a-live-time}

\begin{figure}[h!bp]
    \centering
    \includegraphics[width=0.45\textwidth]{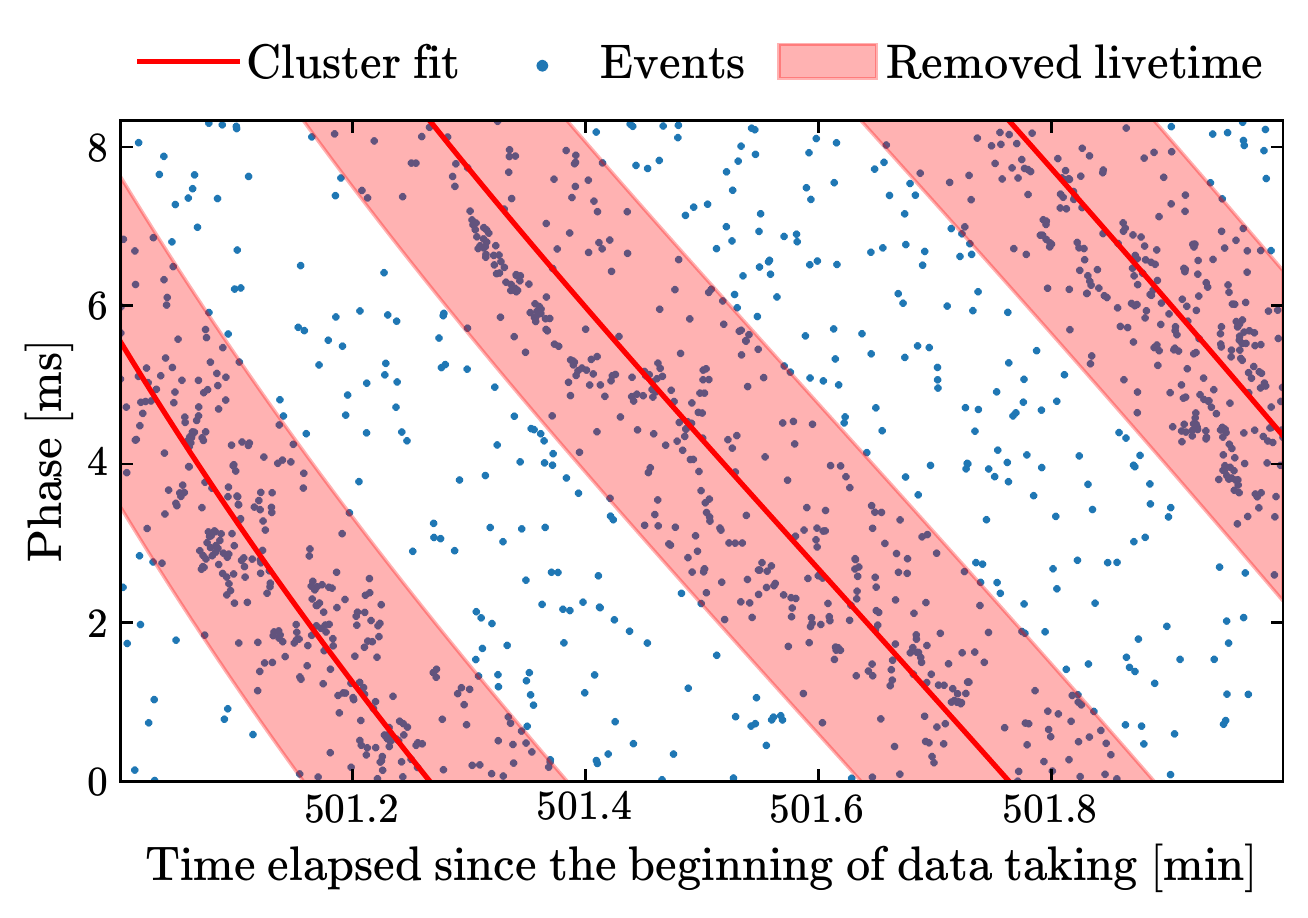}
    \caption{Triggered events in the phase vs. time  plane (blue dots) with the $5^{\mathrm{th}}$-order polynomial fit of the event clusters (red line) and the live-time removed by the 120\,Hz selection (red area). The selection is used to identify and remove periods of high trigger rate associated with power-line-induced noise. The time interval shown in this figure corresponds to approximately 0.3\% of the entire data sample analyzed in this report.}
    \label{fig:120hz_cut}
\end{figure}

We apply the following live-time selection criteria to ensure a stable data-taking environment: (1) a fridge temperature selection discards time intervals with unstable ADR temperatures; (2) an average pre-pulse baseline selection removes time intervals that lie on the tail of high energy particle hits; and (3) a 120\,Hz selection removes time intervals affected by the power-line noise. The selection criteria (1) and (2) are similar to the ones used in the electron recoil dark matter search in Ref.~\cite{Amaral:2020ryn} with the only difference being that we use a time bin of 0.1\,s instead of 1\,s to preserve more live-time. 
The necessity of the 120\,Hz selection (3) arises from the use of a much lower trigger threshold for the 0V data compared to the HV data in Ref.~\cite{Amaral:2020ryn}. We observe that the trigger rate fluctuates with a 120\,Hz frequency. We relate this feature to the power-line-induced noise and identify its phase by clustering triggered events in the phase vs. time plane as shown in Fig.~\ref{fig:120hz_cut}, where the phase is defined as time modulo 1/120\,s. The average phase of the event clusters varies in time due to the varying AC power phase relative to the stable data acquisition clock cycle. We fit the time-dependent phase trend of the increased trigger rate with a $5^{\mathrm{th}}$-order polynomial and remove a 50\% live-time band around the fit, as shown in Fig.~\ref{fig:120hz_cut}. 
After applying all three live-time selection criteria to the 0\,V data, the remaining science exposure is 0.185\,gram·days.

\subsection{Event-based selections}
\label{subsec:4b-eventcuts}

\begin{figure}[htbp]
    \centering
    \includegraphics[width=0.5\textwidth]{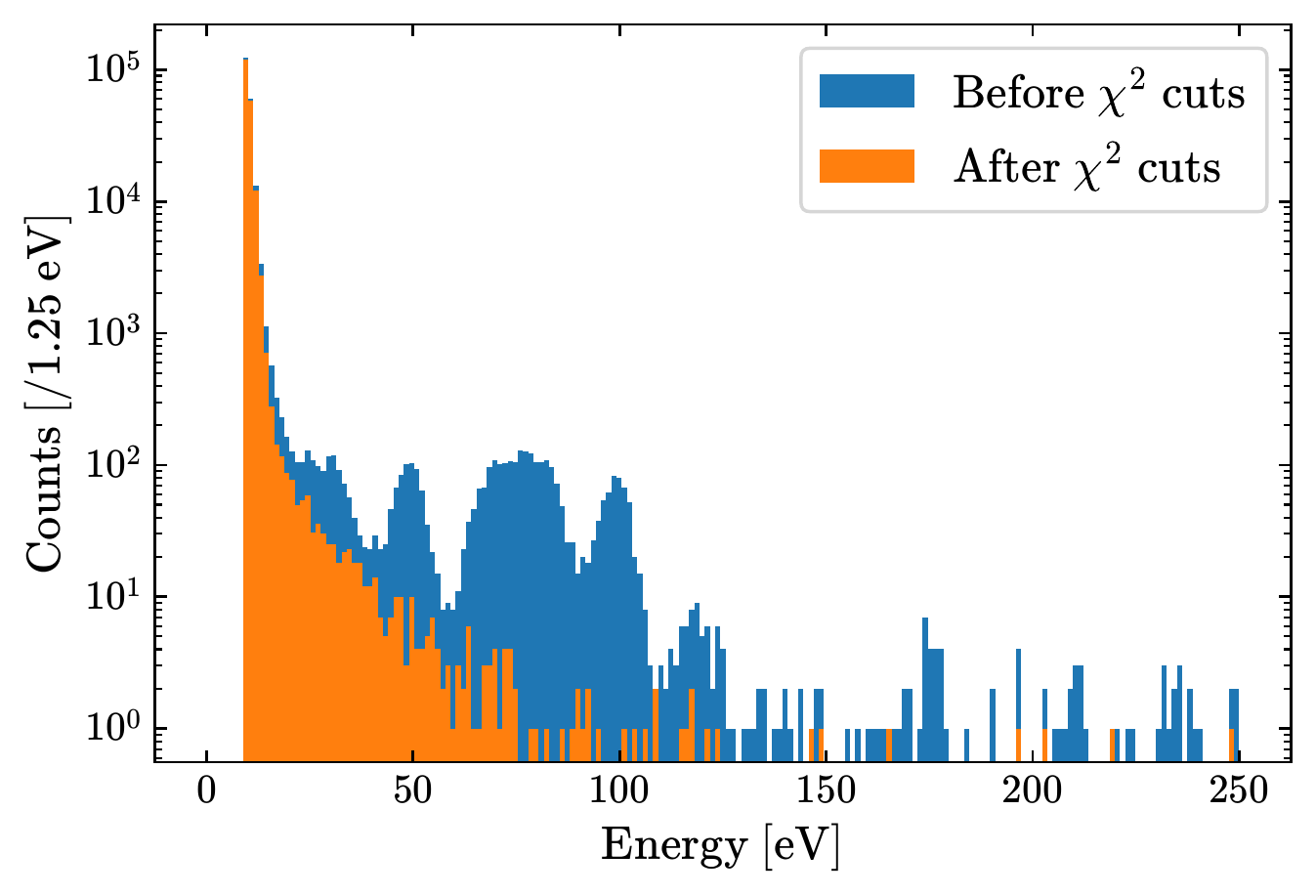}
    \caption{0\,V dark matter search energy spectrum before and after applying the $\chi^2$ selections. The live-time selection criteria are applied to both spectra.}
    \label{fig:0v_spectrum}
\end{figure}

\begin{figure}[htbp]
    \centering
    \includegraphics[width=0.5\textwidth]{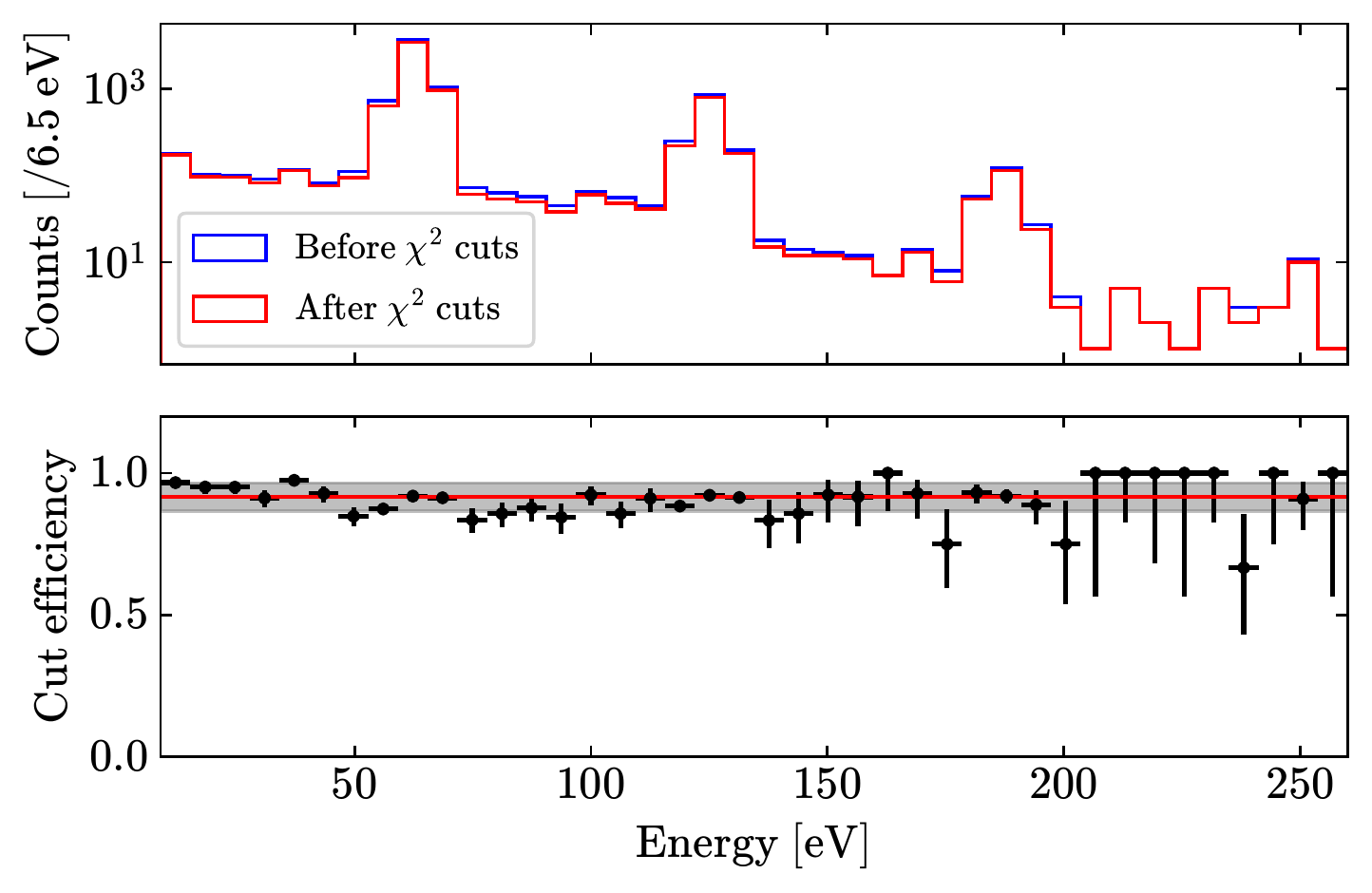}
    \caption{Top: 60\,V laser calibration spectrum before (blue) and after (red) applying the $\chi^2$ selections. Bottom: selection efficiency versus total phonon energy (black data points) fitted by an energy-independent efficiency model (red line) and 1$\sigma$ statistical uncertainty (gray band).}
    \label{fig:cut_efficiency}
\end{figure}

We perform pulse-shape-based selections to remove pulses not consistent with particle energy depositions in the region of interest (ROI) between 9.2 and 250\,eV. The reduced-$\chi^2$, in both the time and frequency domains, between the pulses and the pulse template serves as the metric. We refer to the reduced-$\chi^2$ as $\chi^2$ in this paper for simplicity.
We reject events for which the $\chi^2$ quantity deviates from the corresponding mean of the laser calibration data by over 3$\sigma$, which rejects anomalous triggers such as those caused by electromagnetic interference (EMI) pickup.
Figure~\ref{fig:0v_spectrum} shows the energy spectrum of the dark matter search data before and after applying the $\chi^2$ selections.
The combined efficiency of the two selections is calculated as the passage fraction of the laser data with an energy-independent fit and is shown in Fig.~\ref{fig:cut_efficiency}. We tested how the selection-efficiency uncertainty affects the dark matter limit and found that even a large uncertainty of up to 20\% is subdominant to the other uncertainties, as discussed in the following subsection. Therefore, the $\chi^2$ selection-efficiency uncertainty is not included in the estimate of the systematic uncertainty shown in Fig.~5.

\subsection{Dark matter limit}\label{subsec:4c-limit}

We obtain an exclusion limit on the spin-independent DM-nucleon scattering cross section using a signal-only hypothesis and the data described in the previous subsection.
The calculation uses the standard signal model in Ref.~\cite{ahmed2011search} with the following parameters: an asymptotic value of the Maxwellian velocity distribution $v_0$\,=\,220\,km/s, a galactic escape velocity $v_{\textrm{esc}}$ = 544\,km/s, a local DM mass density $\rho_0$ = 0.3\,GeV/($c^2 \cdot \mathrm{cm}^3$) and a mean orbital velocity of the Earth $v_{\textrm{lab}}$ = 232\,km/s~\cite{Kerr:1986, Smith:2007, Schoenrich:2010}.

To account for the effect of detector resolution on the energy reconstruction, we perform a detector response simulation. We scale the pulse template to energies between 0.5 and 260\,eV in 0.5\,eV steps, and inject these scaled template pulses into randomly triggered noise traces collected throughout the data-taking period. We use the same triggering and energy-reconstruction algorithms that are used for the experimental data to reconstruct the energy of an injected pulse, thus obtaining detector response probability distributions $P(E'|E_0)$, where $E_0$ is the true energy of the injected pulses and $E'$ is the reconstructed energy.
We use a trigger-time selection to ensure that the triggered events correspond to the injected pulses. The dark matter signal model as a function of true energy is then convolved with the detector response probability distributions to construct the signal model as a function of reconstructed energy:
\begin{equation}\label{eqn:sm_convolution}
\begin{split}
    \frac{\partial R}{\partial E'}(E'|M_{\rm DM}) &{}={} \Theta(E' - \delta) \varepsilon \times \\
    \int_{E_0=0{\rm eV}}^{260 {\rm eV}} \bigg[&\Theta\left(E' - E_0 + 3\sigma(E_0)\right)\times\\
    & \Theta\left(E_0 + 3\sigma(E_0) - E'\right)\times\\
    & P(E'|E_0)\frac{\partial R}{\partial E_0}(E_0, M_{\rm DM})\bigg] \mathrm{d}E_0.
\end{split}
\end{equation}
Here $E_0$ is the true recoil energy, $E'$ is the reconstructed energy, $\frac{\partial R}{\partial E_0}$ is the differential DM-nucleon scattering rate, $M_{\rm DM}$ is the dark matter candidate mass, $\delta$ is the trigger threshold, and $\varepsilon$ is the selection efficiency (assumed energy-independent in this analysis). The trigger efficiency is included in the detector response probability distributions $P(E'|E_0)$. The two Heaviside functions $\Theta$ inside the integral perform a 3$\sigma$ cutoff of the detector response function, where $\sigma(E_0)$ is the width of the Gaussian fits to each $P(E'|E_0)$ distribution. This cutoff simplifies the numerical calculation by restricting the convolution of the detector response with the signal model to a range of $\sim$ 1.7\,eV to 258.7\,eV and avoids an undefined recoil rate at zero energy.

\begin{figure*}[htbp]
    \centering
    \includegraphics[width=0.98\textwidth]{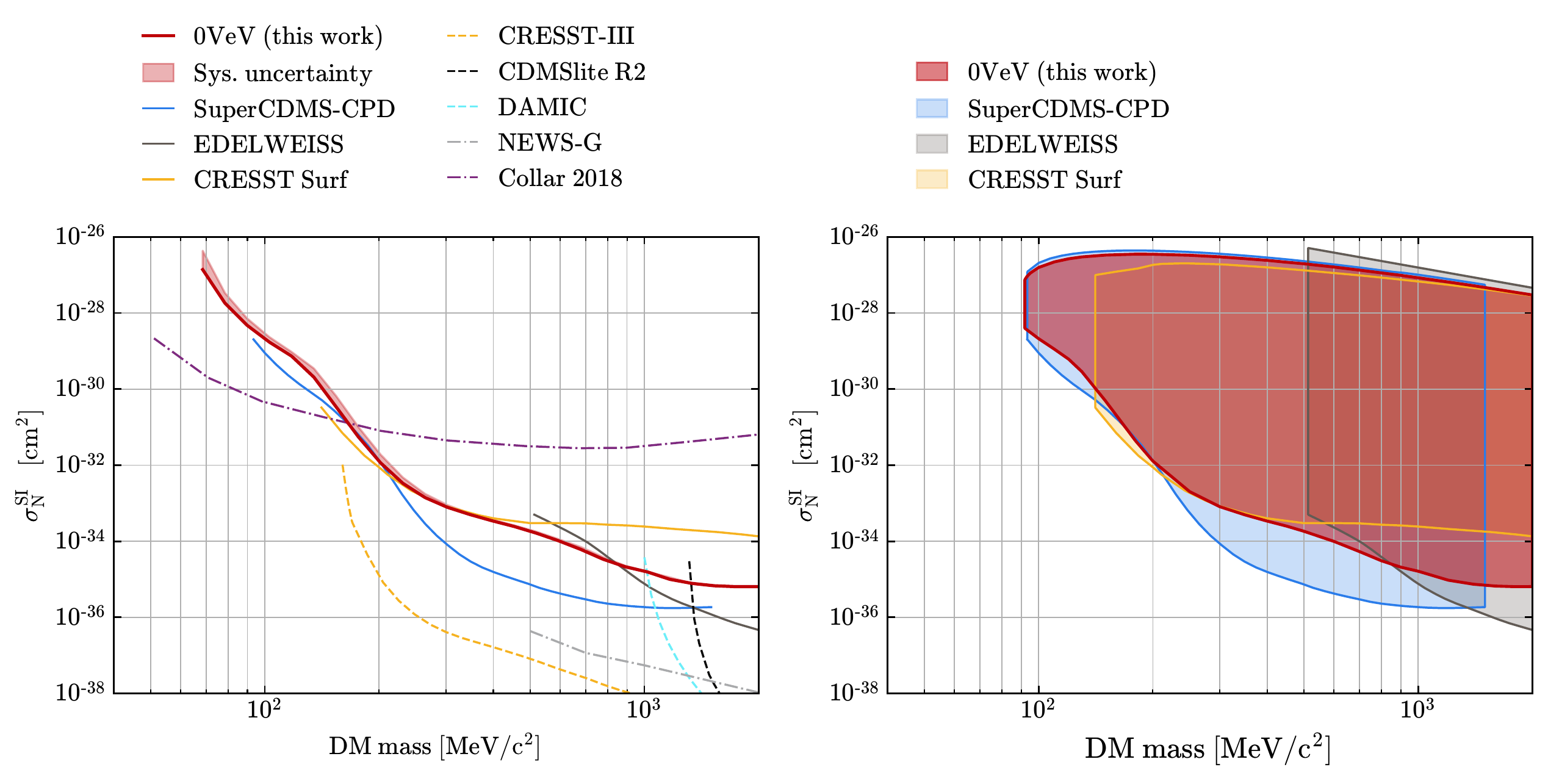}
    \caption{Left: 90\% confidence-level exclusion limit on the spin-independent DM-nucleon scattering cross section. The result of this work is depicted in solid red with an estimate of the systematic uncertainties in light red. The results of other surface experiments using solid-state detectors are depicted as solid lines: blue for SuperCDMS-CPD \cite{Alkhatib:2020cpd}, dark gray for EDELWEISS \cite{Armengaud:2019}, and gold for CRESST-surface \cite{Angloher:2017}. Underground searches  using solid-state detectors are depicted as dashed lines: gold for CRESST-III \cite{Abdelhameed:2019}, dark gray for CDMSlite \cite{Agnese:2016}, and cyan for DAMIC \cite{Aguilar-Arevalo:2016}. Other experimental constraints are shown as dash-dotted lines: light gray for NEWS-G \cite{Arnaud:2017bjh} and purple for Collar \cite{Collar:2018}. Right: the same results with upper- and low-mass boundaries on the exclusion areas derived from the atmosphere and Earth shielding effect~\cite{Alkhatib:2020cpd,Armengaud:2019,Davis:2017}. The upper boundary limits the low-mass reach of the current experiment to 92\,MeV/$c^2$.}
    \label{fig:dm_limit}
\end{figure*}

We utilize the Optimum Interval (OI) method~\cite{Yellin:2002,Yellin:2007} to set a 90\% confidence level exclusion limit on the DM-nucleon scattering cross section, using the experimental spectrum and the signal model described above. Figure~\ref{fig:dm_limit}~(left) compares our result to other experimental results in the same mass region~\cite{Alkhatib:2020cpd,Armengaud:2019,Angloher:2017,Abdelhameed:2019,Agnese:2016,Aguilar-Arevalo:2016,Arnaud:2017bjh,Collar:2018}. The systematic uncertainty propagated from the energy calibration uncertainty, discussed in Sec.~III~B, is shown as the filled area. We estimate the systematic uncertainty by rescaling the energy calibration by 11\% (see Fig.~\ref{fig:calibration}) and recalculating the limit. The resulting limit differs from the main result by up to 6$\times$ at the lowest mass (up to 2$\times$ at masses above 100\,MeV/$c^2$). The other systematic uncertainties are not included in Fig.~5 as they were found to be subdominant: up to 20\% from the uncertainties in the detector response simulation and less than 20\% from the cut-efficiency uncertainty.

A very-low-energy threshold allows us to reach dark matter masses below 100\,MeV/$c^2$, but the relatively high cross-section values in this mass range require us to consider the shielding by the atmosphere and Earth. 
At high values of the cross section, a presumed dark matter particle would not reach the detector due to its interactions with the atmosphere and the Earth, therefore such cross-section values cannot be probed by our experiment.
To calculate the upper bound on the cross-section exclusion region (Fig.~\ref{fig:dm_limit}, right), we use the \texttt{verne} package~\cite{Kavanagh:2017}, which takes into account the mean direction of the DM flux at the location and the time of the experiment and estimates the impact of shielding on the standard halo model velocity distribution, assuming straight-line particle trajectories and continuous energy loss in the shielding (atmosphere and Earth). While these assumptions are in general only valid for high-mass particles ($>10^5\,\mathrm{GeV}/c^2$), a comparison with a more complete Monte Carlo approach demonstrates that the simplified approach used in the \texttt{verne} package leads to similar results~\cite{Emken:2018}.  Accounting for shielding removes the sensitivity of this analysis to dark matter masses below 92\,MeV/$c^2$. To make a comparison to other experimental results in the same parameter space~\cite{Alkhatib:2020cpd,Armengaud:2019,Davis:2017}, we do not correct the lower bound of the exclusion region for shielding. However, this correction should be done in general at cross sections $\gtrsim10^{-33}$\,cm$^2$, especially for experiments probing new parameter spaces. Further efforts are required to consider shielding in the OI method, as it introduces a dependency of the DM spectrum shape on the value of the cross-section. In the current analysis, if the entire energy ROI is used instead of the OI method, considering DM shielding would increase the lower bound of the exclusion region by a factor of $\sim$2.1 at 100\,MeV/$c^2$.

\section{Pulse shape anomalies}\label{sec:5-pulse_shape}

\begin{figure}[htbp]
    \centering
    \includegraphics[width=0.47\textwidth]{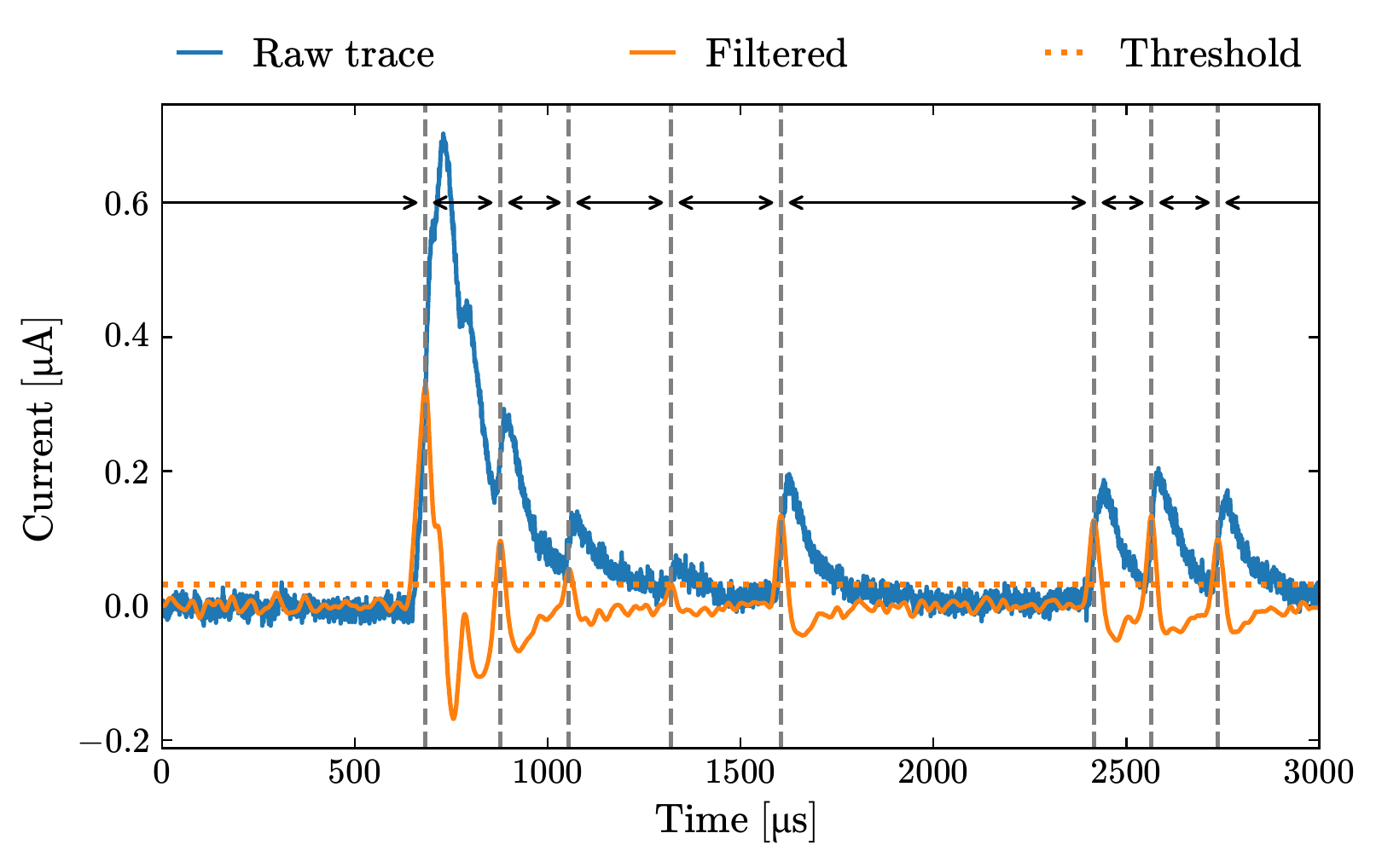}
    \caption{Example of a ``burst" event at 60\,V. The blue trace is the raw trace,  whereas the orange trace results after applying a Gaussian derivative filter (described in Sec.~\ref{subsec:HVmode:burstsEvents}), which peaks at the rising edges in the raw trace. The dotted orange line is the threshold for peak-finding. Each peak above the threshold in the filtered trace corresponds to a pulse in the raw trace. Note that the filter has limited time resolution, which results in the second pulse being below the threshold and not identified. The vertical dashed guide lines show the rising edge of the events identified above the threshold.
    The inter-arrival time of two events is defined as the time distance between their rising edge. 
    }
    \label{fig:burst_demo}
\end{figure}

We observe populations of events with pulse shapes different from the calibration data in the data-set even after the $\chi^2$ cut. Anomalously shaped events exist in both the 0V and HV DM exposures with different characteristics. In the 0V data, we observe events that have a significantly longer pulse decay time than the laser-pulse shape. In HV data, we notice a large population of events with more than one pulse closely packed in time, which we refer to as ``burst" events in this manuscript. Figure~\ref{fig:burst_demo} shows one example of a burst event. 
To study these anomalous events, we do not use the event-based selections described in Sec.~\ref{sec:4-DM} because they tend to remove these events. We instead establish looser selections described in this section and use them to investigate the pulse shape anomalies in the 0V and HV data. We then discuss the pulse shape anomalies in 0V and HV data in the rest of this section.

\subsection{Data selection\label{subsec:0vhvcuts}}
To study the pulse shape anomalies and facilitate the comparison of the 0V and HV datasets, we apply the same live-time selections (1) and (2) described in Sec.~\ref{sec:4-DM} to both datasets. We increase the analysis threshold for this investigation to 25\,eV to avoid near-threshold noise effects such as the the 120\,Hz power-line-induced noise events, which allows us to preserve more exposure because live-time selection (3) is not needed. The resulting exposures are 0.4~gram·days at 0\,V, 0.7~gram·days at 60\,V, and 1.7~gram·days at 100\,V.

We use a loose $\chi^2$ selection to remove trigger artifacts caused by the OF. We also use a pulse-width selection to reject EMI noise, for which the average pulse width is wider ($>$\,\SI{160}{\micro\second})  than for particle-interaction events ($<$\,\SI{100}{\micro\second}). The two selections are applied to both HV and 0V data. The selection efficiencies are evaluated in Sec.~\ref{sec:5-pulse_shape}. 

For the pulse-shape study reported in this section, we also remove a population of ``slow events" from the 0V data. These events have pulse-decay times two orders of magnitude slower than the decay time for laser-calibration events. Such a slow time constant indicates that these events are the result of a different type of energy deposition in the detector. We discuss this class of events further in Sec.~\ref{sec:discussion} B.

\subsection{0V mode: long-tail events}
\begin{figure}
    \centering
    \includegraphics[width=0.45\textwidth]{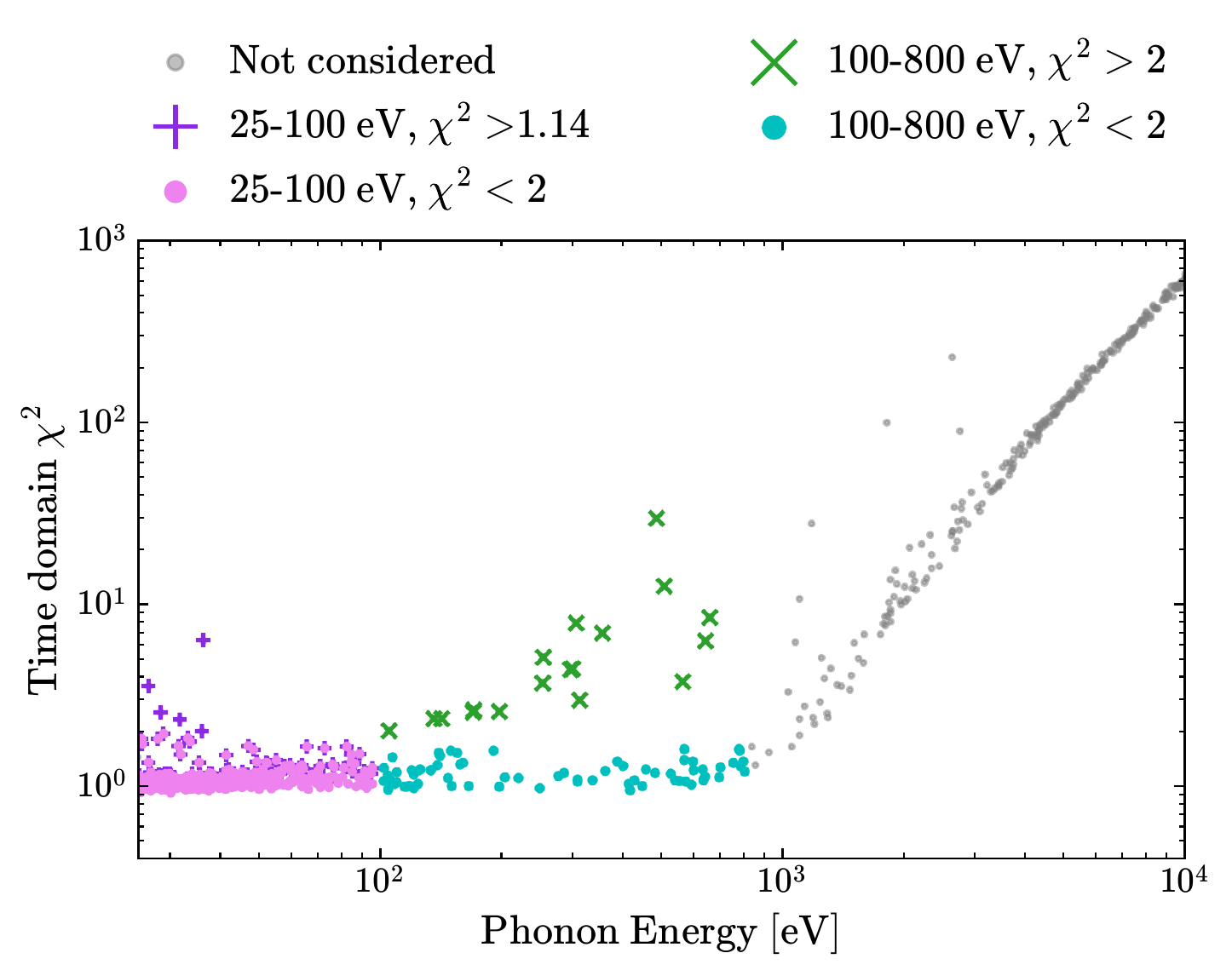}
    \includegraphics[width=0.45\textwidth]{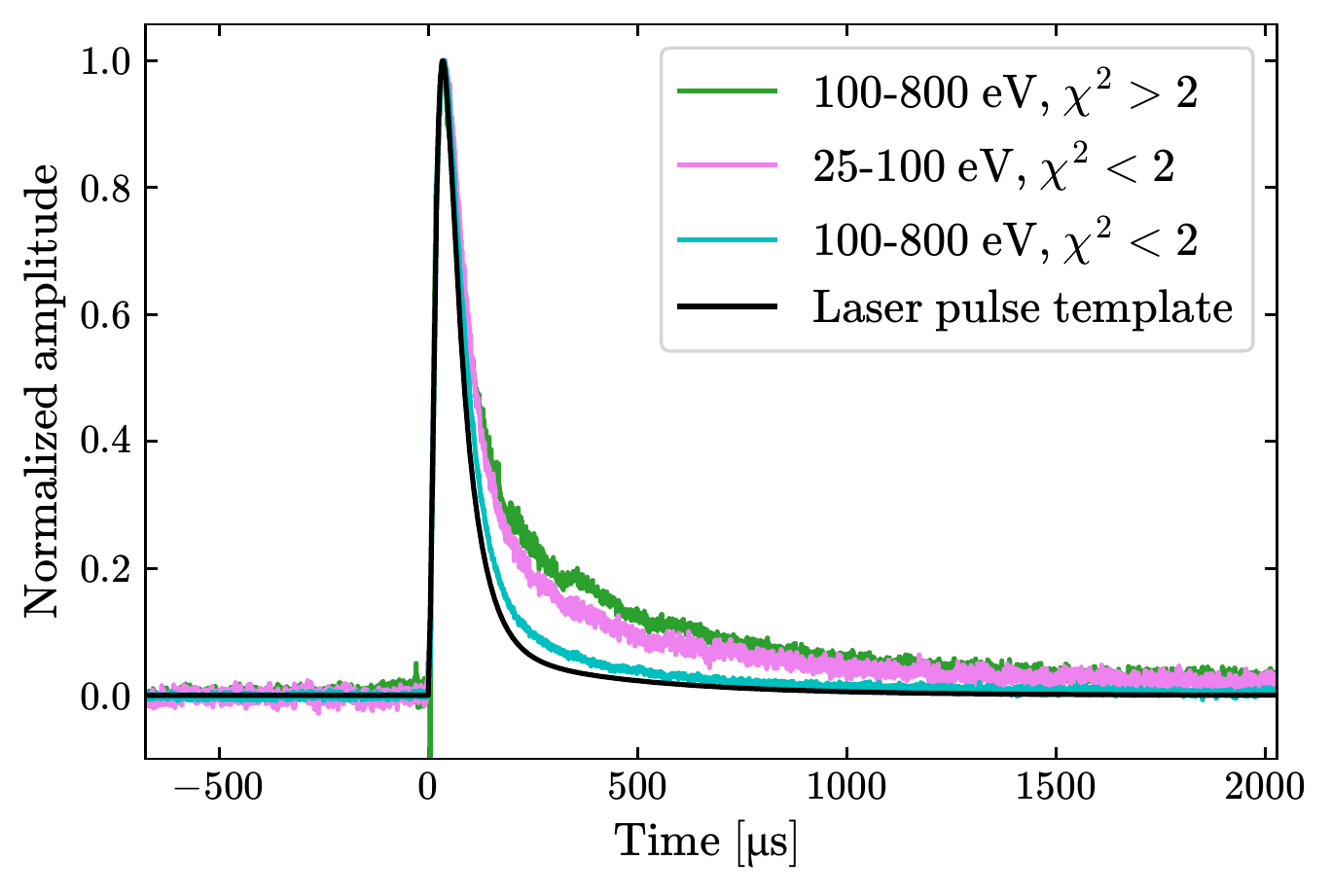}
    \caption{Event selection (top) of 0V data and averaged pulses of selected 0V events (bottom). Four groups of events are selected with two energy ranges (25--100 eV and 100--800 eV; events $>$800 eV, shown in gray, are not considered due to detector saturation) and two $\chi^2$ ranges ($\chi^2>2$ and $\chi^2<2$). The $25-100$\,eV events with $\chi^2>2$ (marked with purple +) have square pulse shape which is consistent with Radio Frequency (RF) induced noise, and are not plotted in the bottom panel. The legend in the bottom panel is ordered from top to bottom with decreasing fall time of the averaged pulse shape.}
    \label{fig:longtail_selection}
\end{figure}

The $\chi^2$ metric is sensitive to differences in pulse shape relative to the pulse template, and different event populations are apparent in the $\chi^2$ versus reconstructed-energy plane (Fig.~\ref{fig:longtail_selection} top) for the 0V data. Using event selections in this plane, we create average pulses for each group (Fig.~\ref{fig:longtail_selection} bottom). We split the data into a low-energy region (up to 100\,eV) where the signal-to-noise ratio is modest and a high-energy region from 100--800 eV where pulse-shape differences are more easily distinguishable by $\chi^2$.
Events above 800\,eV are subject to strong detector saturation effects and have hence been excluded in this pulse-shape study. 
For each energy region, we select events with a template-like shape with an empirical selection of $\chi^2 < 2$ and an anomalous shape with $\chi^2 > 2$.
We compare these to the aforementioned template made with laser pulses. To rule out pulse-shape differences associated with different interaction types, we verified that this pulse template is also consistent with the pulse shape of nuclear recoil events both at 0\,V and 100\,V, using data taken at a neutron beam~\cite{CDMS:2021}.

The average pulse of the anomalous $\chi^2>2$  events between 100\,eV and 800\,eV, shown in green in Fig.~\ref{fig:longtail_selection}\,(bottom), exhibits a pronounced slower decay time, or ``long tail", compared to the pulse template. The average pulse of events in this energy range with $\chi^2 < 2$ is very similar in shape to the pulse template, see the cyan pulse in Fig.~\ref{fig:longtail_selection}\,(bottom). The small deviation of the 100-800\,eV average pulse (cyan) from the template is a result of including some events with slight saturation and some of the long-tail events. As is visible in Fig.~\ref{fig:longtail_selection}\,(top) the discrepancy in $\chi^2$ diminishes with decreasing energy and is close to our selection boundary at $\sim 100$\,eV. Hence, we do not expect a full event-by-event separation of these long-tail events for the low-energy selection of $\chi^2<2$ events (in pink). Curiously, we observe an average pulse from this population that is much closer to the pulse shape of the anomalous events in the 100--800\,eV range than that of the laser-pulse template. This suggests that the low-energy data are dominated by long-tail events.

\subsection{HV mode: burst events}
\label{subsec:HVmode:burstsEvents}
When the detector is operated in HV mode, we classify all events with more than one pulse in the 5.4 ms post-trigger time window as a burst event, as exemplified by the event shown in Fig.~\ref{fig:burst_demo}.
We divide the pulses in a burst event into two categories: the primary pulse occurring at the trigger time of the event, and the secondary pulses occurring after the primary pulse. 
Pulses from both categories are treated as a single event.

\begin{figure}[hpbt]
    \centering
    \includegraphics[width=0.45\textwidth]{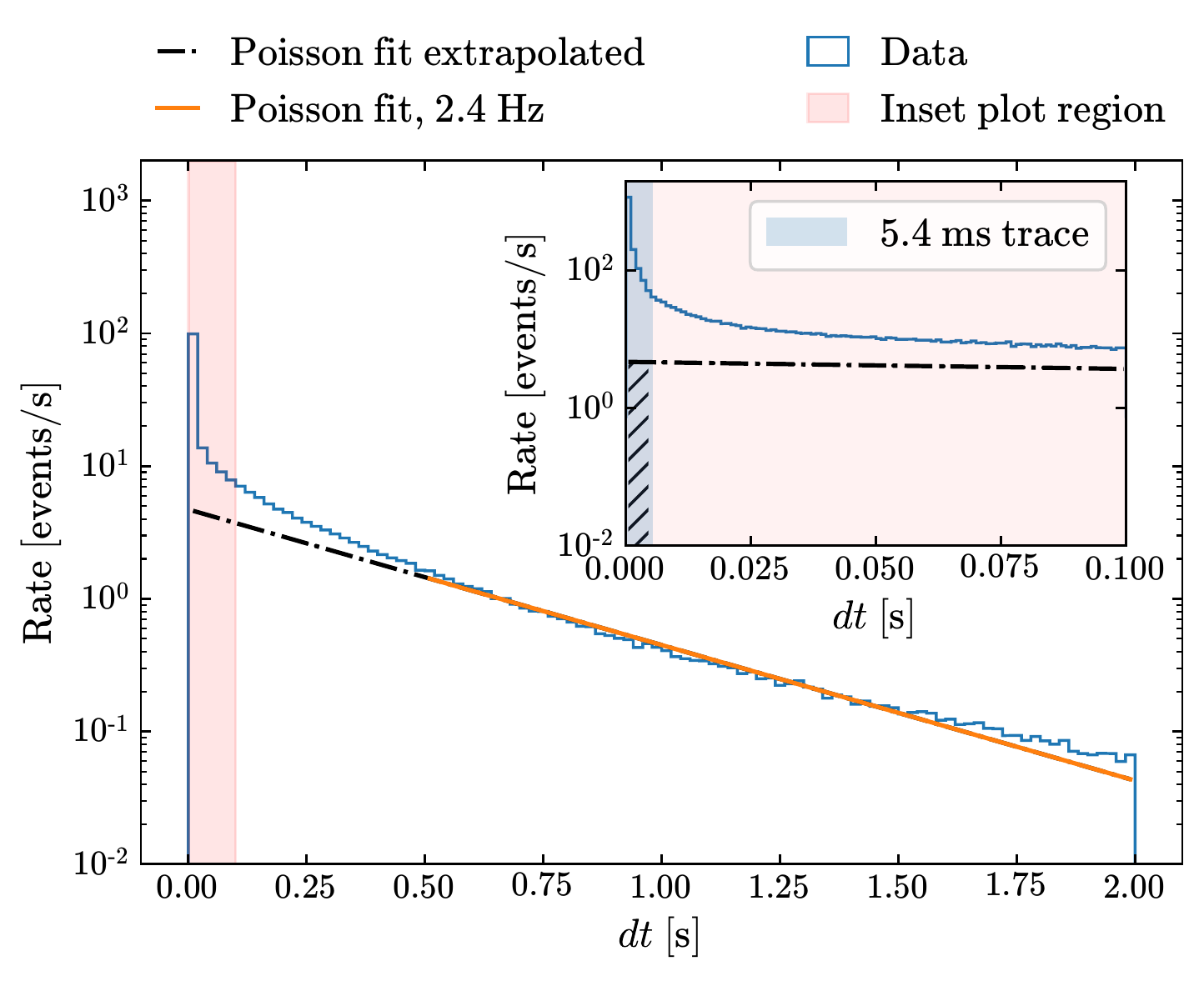}
    \caption{Individual-pulse inter-arrival time ($dt$) distribution of 100\,V data. The bin width of the main plot is 0.02s. The inset panel is a zoom-in of the highlighted pink region. The blue area in the inset plot indicates the post-trigger duration used in our standard event-reconstruction algorithm.}
    \label{fig:dt}
\end{figure}

To study the time distribution of the individual pulses, we identify the individual pulses inside a burst event with an edge detection algorithm. This algorithm searches for peaks after filtering the raw event with a first-order Gaussian derivative kernel. 
The inter-arrival time ($dt$) is defined as the time distance between sequential rising edges as shown in Fig.~\ref{fig:burst_demo}. The $dt$ distribution of all pulses is shown in Fig.~\ref{fig:dt}. 
If all the pulses were from a random Poissonian process with uncorrelated pile-up probabilities, the $dt$ distribution would follow a single exponential function. We note that the distribution roughly follows such an exponential function in the region of $0.5\ \mathrm s<dt<1.5\ \mathrm s$, while deviating from it at smaller and larger time scales. The deviation at larger time scales suggests there may be long-time correlation between events, though this is not investigated in this report. Meanwhile, at smaller time scales the non-Poissonian component dominates. For example, within the post-trigger trace length of 5.4\,ms, the Poissonian component contributes only 2\% of all pulses. This suggests that the majority of the individual pulses in burst events are correlated in time and likely have a common origin.

\begin{figure}[htbp]
    \centering
    \includegraphics[width=0.45\textwidth]{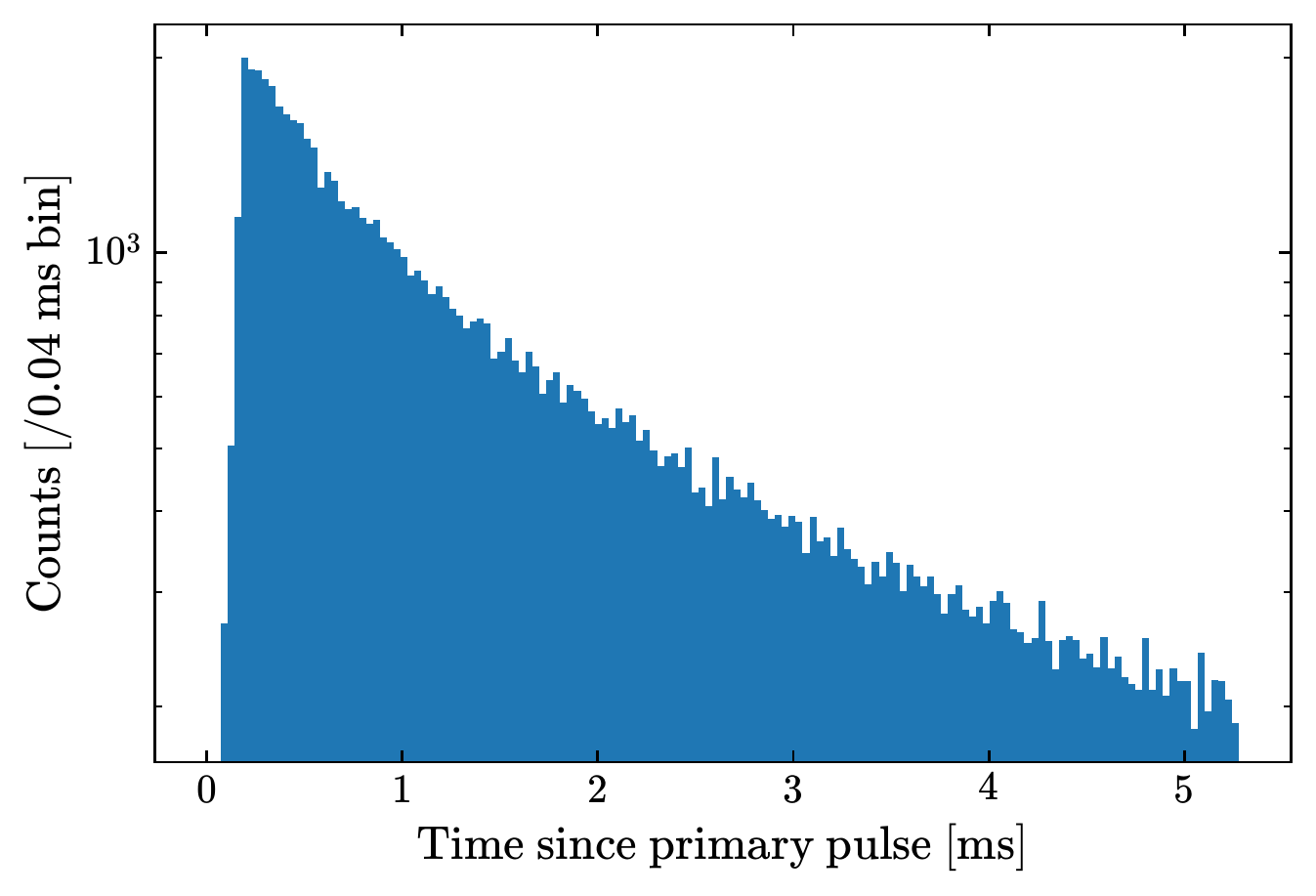}
    \caption{Time distribution of secondary pulses with respect to the primary (triggering) pulse in burst events from HV-mode data with the detector operated at 100\,V. Note that the first five time bins (starting at 0 ms) have few counts due to the limited time resolution of the peak finding algorithm.}
    \label{fig:burst_secondary_time}
\end{figure}

We further characterize the burst events via the distribution of secondary-pulse arrival times relative to the primary pulse, Fig.~\ref{fig:burst_secondary_time}. This time distribution is used later in Sec.~\ref{sec:6-compare}~B to simulate burst events.
The rate of secondary pulses decreases non-exponentially, which suggest there are multiple time scales.

The high rate of secondary pulses within a short time requires a special methodology to reconstruct their individual energies. First, we use a much shorter trace length of $\sim$\SI{150}{\micro\second} as opposed to the 10.8 ms used in our standard event reconstruction. We then fit the pre-pulse baseline with an exponential function and subtract this function from the trace to minimize the impact of the preceding pulse on the reconstructed energy. Finally, we correct for the baseline-dependent gain variations as defined in Ref.~\cite{ren:2020} and use the best-fit OF amplitude to estimate the energy.

\begin{figure}[htbp]
    \centering
    \includegraphics[width=0.45\textwidth]{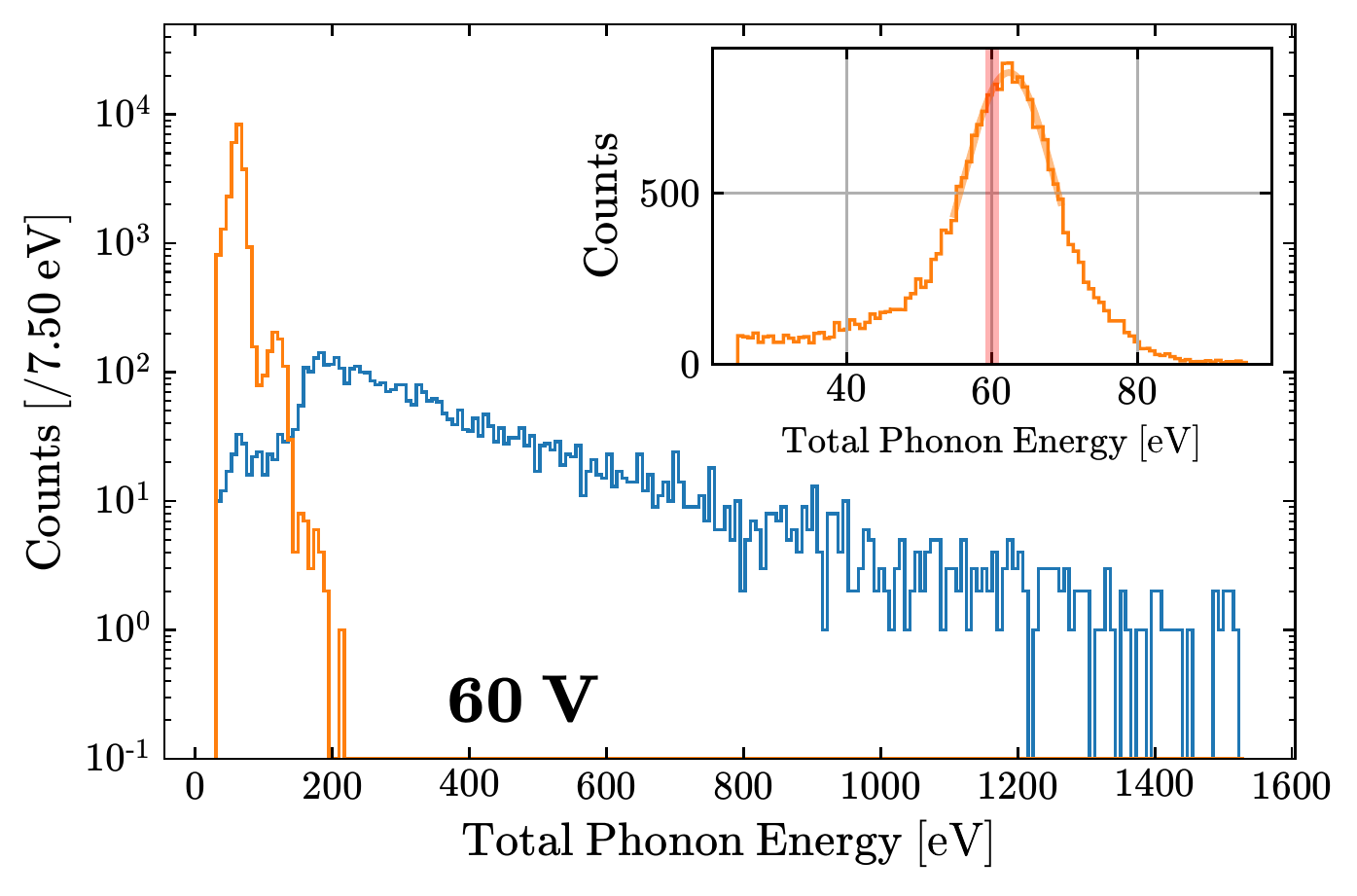}
    \includegraphics[width=0.45\textwidth]{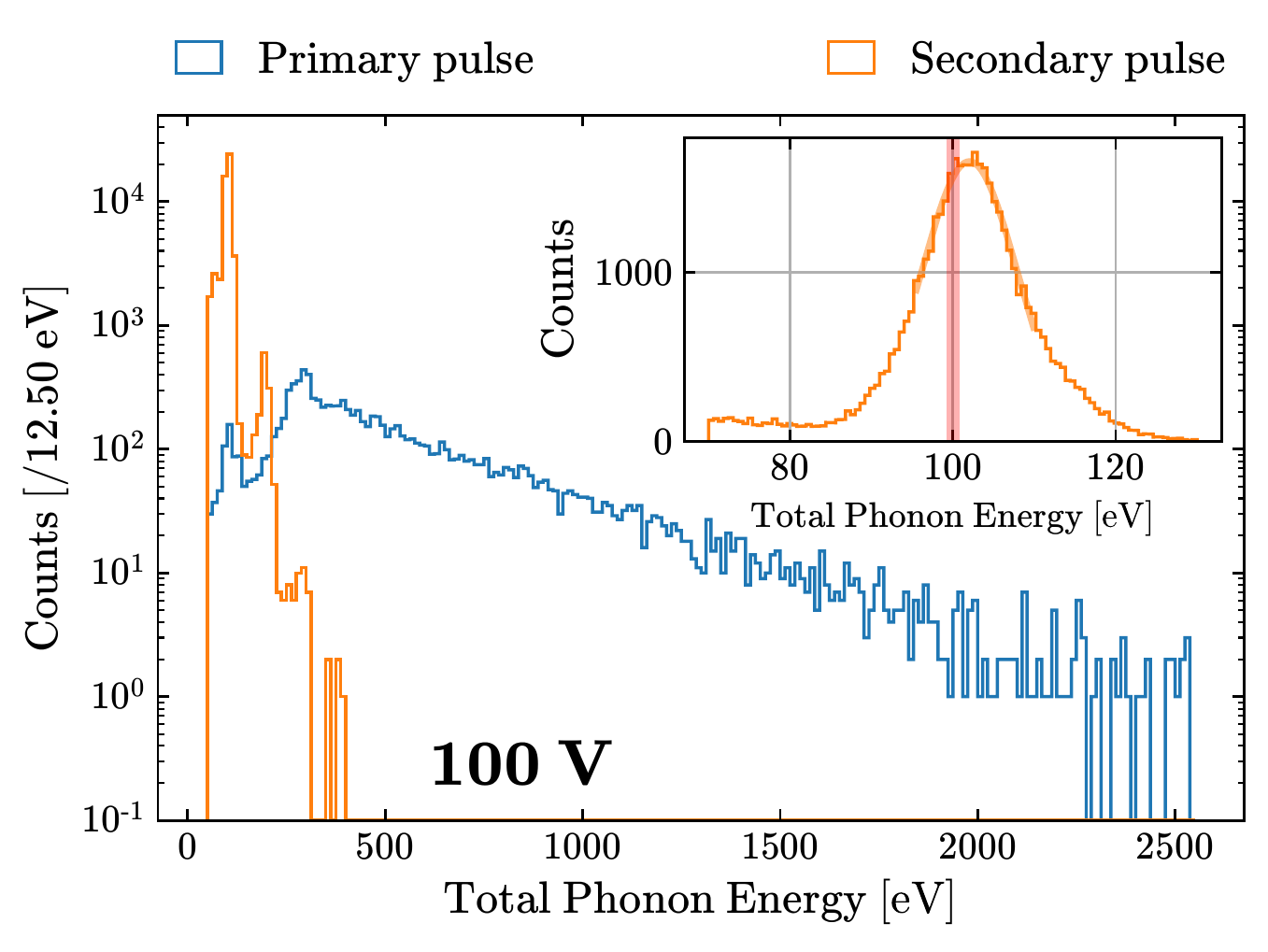}
    \caption{Energy of primary and secondary pulses of burst events in  60\,V and 100\,V  data. The first peak in the primary pulse spectrum is consistent with 60\,eV and 100\,eV, respectively. The inset plot shows the zoom-in of around the energy of 1 \eh for secondary pulses. The red vertical line in the inset plot indicates the energy of the NTL contribution ($e\cdot \mathrm{V_{\mathrm{NTL}}}$) for a single \eh, which is 60\,eV and 100\,eV, respectively. }
    \label{fig:burst_energy}
\end{figure}

The energy spectra of the primary pulses and the secondary pulses are shown in Fig.~\ref{fig:burst_energy}. We note that the primary pulse energy goes up to several keV, while the secondary pulse energy peaks around the energy of a single \eh. The energy of single \eh events is given by the initial recoil energy $E_{\mathrm{r}}$ and the NTL phonon energy, $e\cdot V_{\mathrm{NTL}}$. 
The distribution of secondary pulses peaks at $\sim$2\,eV above $e\cdot V_{\mathrm{NTL}}$; this excess is interpreted as the recoil energy, where the systematic uncertainty of the energy calibration is estimated to $\mathcal{O}(1)$\,eV. 

\section{Comparison between 0V and HV}\label{sec:6-compare}

The only difference between the 0V and HV datasets is the crystal voltage bias; so, we consider the possibility that the anomalous pulse shapes in the 0V data have the same origin as the burst events in the HV data. 
Under this assumption, we compare the 0V and HV pulse shapes based on ensemble averages which will be done in Sec.~\ref{sec:6-compare}~A. In order to also make a spectral comparison and take into account potential effects of the event selections and detector response, we develop a burst event simulation to estimate the detector response for burst events with and without NTL amplification. The simulation is described in Sec.~\ref{sec:6-compare}~B.
Note that while we expect a nonzero voltage bias to introduce charge-leakage events in the HV data that will not be present in the 0V data, these events are below the energy region of interest for the comparison discussed in this section.
We also note that we cannot rule out the alternative hypothesis that the crystal voltage bias can induce time correlated events. We will elaborate on this point in Sec.~\ref{sec:discussion}.

\subsection{Pulse shape comparison}

\begin{figure}[hbpt]
    \centering
    \includegraphics[width=0.45\textwidth]{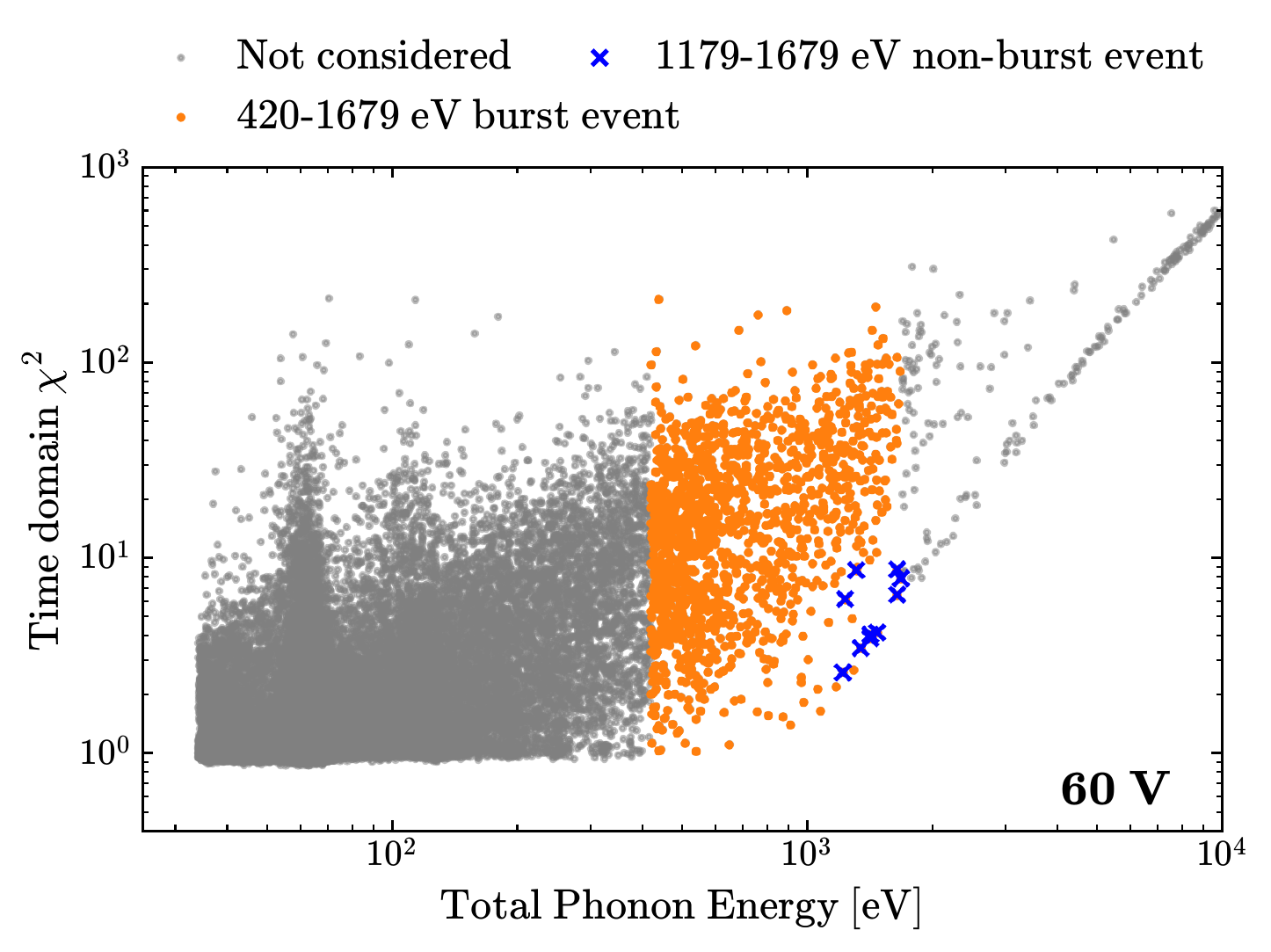}
    \caption{Selection of  HV events for the comparison with 0V long-tail events. All events within 420-1679\,eV and non-burst events at the higher end of that energy range are highlighted in orange dots and blue crosses, respectively.}
    \label{fig:burst_60V_selection}
\end{figure}

\begin{figure}[htpb]
    \centering
    \includegraphics[width=0.45\textwidth]{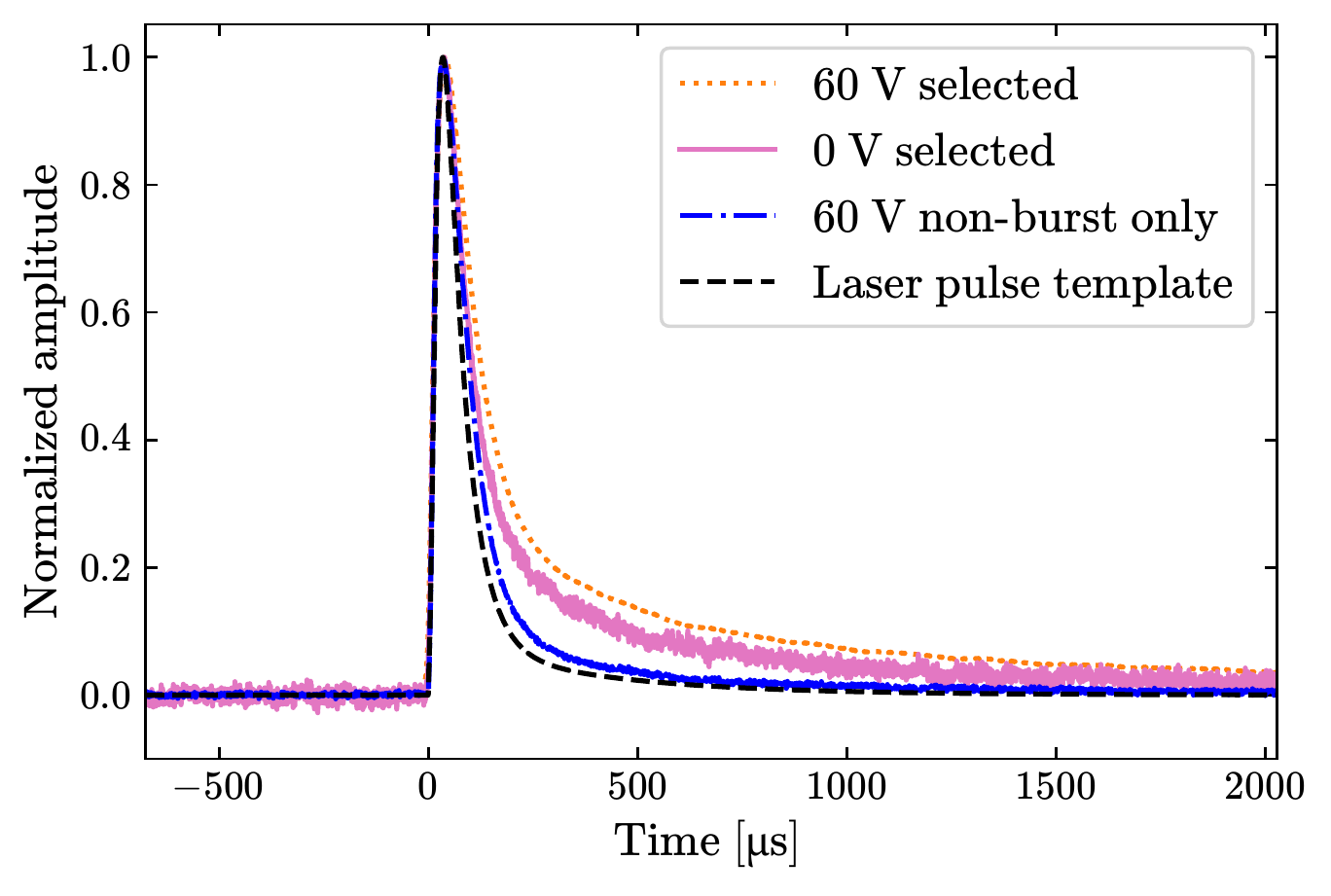}
    \caption{0V vs. HV pulse-shape comparison. Pink and orange-dotted lines are the the average pulse shapes for the 0\,V and 60\,V events selected in Fig.~\ref{fig:longtail_selection} and \ref{fig:burst_60V_selection}. The black-dashed line is the laser-pulse template, which represents the non-saturated pulse shape. The blue line is the average pulse shape for the 60\,V non-burst event sample and acts as a reference of the slightly saturated pulse shape.}
    \label{fig:HV-0V_pulseshape}
\end{figure}

At 0\,V, we cannot distinguish events with an energy that would typically produce a single \eh  from random noise fluctuations, making it difficult to identify potential burst events at 0\,V. Thus, we focus on the averaged pulse shape when comparing between the 0V and HV data. We select 0V data in the energy range between 25\,eV and 100\,eV (pink events in Fig.~\ref{fig:longtail_selection}). The 60\,V events shown as orange dots in the $\chi^2$ vs. energy plane in Fig.~\ref{fig:burst_60V_selection} are chosen to match this energy range with an NTL gain of 16.8, assuming $\eeff=3.8$\,eV~\cite{Ogburn:2008}. Additionally, a subset of 60\,V events that are not burst events (blue crosses in Fig.~\ref{fig:burst_60V_selection}) are also selected at the higher end of this energy range, which have no more than one pulse identified within the 5.4\,ms post-trigger window and thus are less likely to be burst events. We use this group of ``non-burst" events from the HV data to produce an average pulse shape for events that have some saturation. The resulting averaged pulse shapes are shown in Fig.~\ref{fig:HV-0V_pulseshape}.

The average pulse shapes for both the 0V sample and the HV data burst events show visibly longer decay times than the laser-pulse template, which suggests the potential for these 0V and HV events to have a common origin. Conversely, the average pulse shape of the non-burst HV sample is similar to the laser-pulse template, indicating that detector saturation effects are unlikely to be the cause of the longer decays times in the other samples.

\subsection{Burst event simulation}\label{sec:burst_simulation}

The different energy estimators---OF amplitude in the low-energy region, and MF integral in the high-energy region---have different sensitivities to secondary pulses, which is expected to lead to a systematic bias when scaling the HV-mode spectra for comparison to the 0V spectrum.
We correct for this bias by applying a response matrix evaluated with the burst event simulations described below. We also use the burst event simulation to evaluate the event selection efficiencies.

We simulate the burst events with the time and energy distributions measured in the 60\,V dataset. Burst events are characterized by the following parameters:
\begin{itemize}
  \item Primary-pulse energy, $E_\mathrm{p}$
  \item Number of secondary pulses, $N_\mathrm{s}$
  \item Energy of the secondary pulses, $E_\mathrm{s}$
  \item Time of each secondary pulse, $t_\mathrm{s,i}$.
\end{itemize}

\begin{figure}[hbpt]
    \centering
    \includegraphics[width=0.45\textwidth]{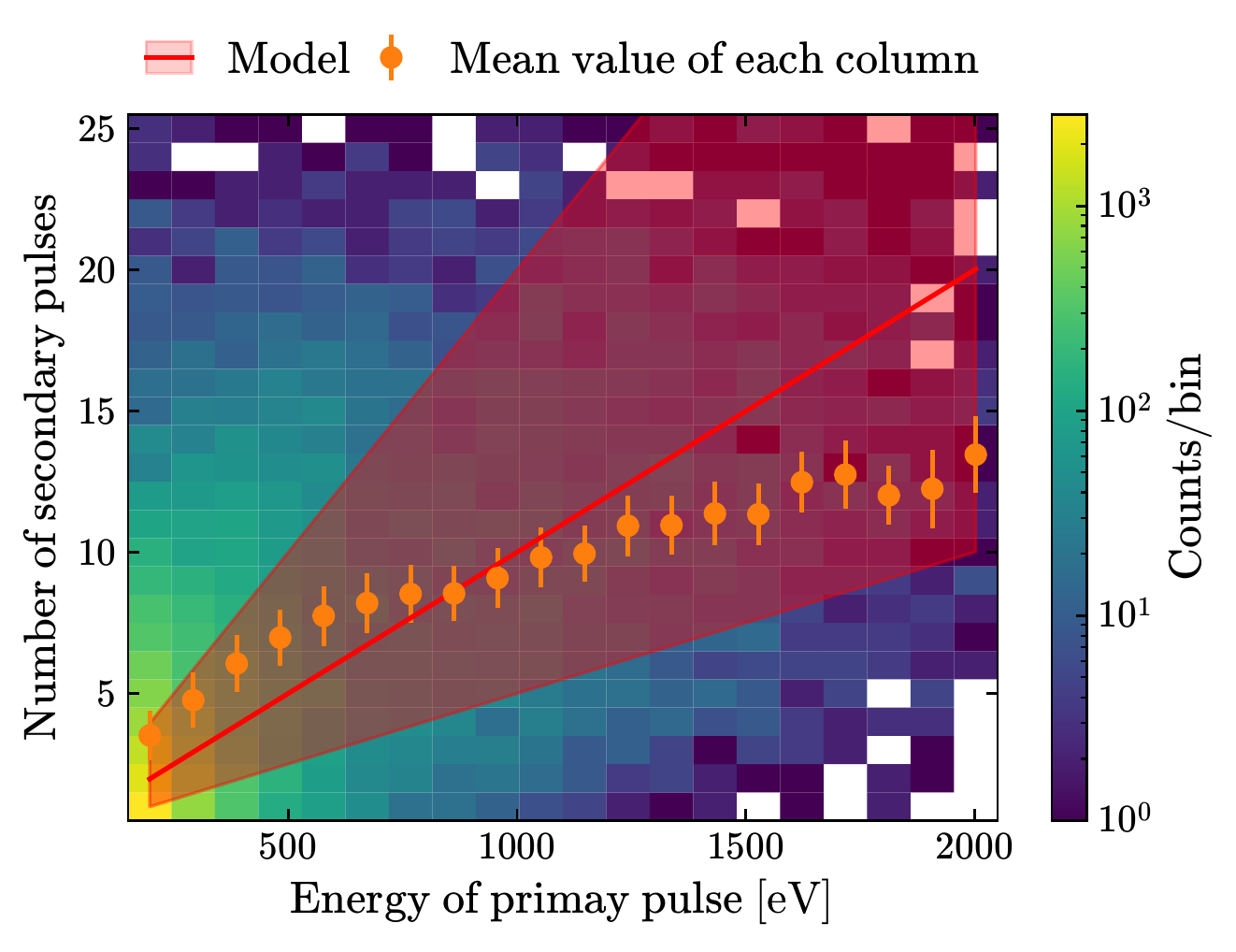}
    \caption{2D histogram of the number of secondary pulses $N_\mathrm{s}$ as function of the primary pulse energy. Orange dots with error bar are the mean and standard deviation of each column of the histogram. The red line represents the relation between the number of secondary pulses and the energy of the primary pulse used in the burst event simulations. The boundaries of the red shaded area, corresponding to double and half the number of secondary pulses relative to the red line, are also simulated.}
    \label{fig:N_pileup}
\end{figure}

We modeled the distributions of $E_\mathrm{p}$ and $t_\mathrm{s,i}$ with probability density functions extracted from the data, conforming to the distribution shown Fig.~\ref{fig:burst_secondary_time} and ~\ref{fig:burst_energy}, respectively. $E_\mathrm{s}$ is set to 2\,eV, which is consistent with single \eh events. 
The distribution of $N_\mathrm{s}$ from data is shown in Fig.~\ref{fig:N_pileup}, and is modeled as a linear function of the energy of the primary pulse with a Gaussian distribution and standard deviation equal to its mean value, as a trial ansatz. The model with nominal parameters is shown as the center red line. The boundaries of the red shaded region, corresponding to double and half the number of secondary pulses compared to the red line, are chosen to bracket the mean number of secondary pulses we observed in data. We simulated three different scenarios corresponding to the red line and the upper and lower edges of the red shading region in Fig.~\ref{fig:N_pileup}. 

We construct the trace of each event by summing a noise trace obtained from randomly triggered data, a primary pulse with the energy-dependent pulse shape empirically determined from calibration data, and $N_\mathrm{s}$ secondary pulses using the pulse template and onset times following the $t_\mathrm{s,i}$ distribution. The simulated data sets are then reconstructed using the same algorithms as the detector data.

\subsection{Energy spectra comparison}

The energy spectra measured with a crystal voltage bias of 60\,V and 100\,V correspond to the total phonon energy with NTL gain, while the energy spectrum measured at 0\,V represents the recoil energy. The NTL gain depends on the averaged \eh production energy, $\eeff$. By comparing the spectra at different voltages we can estimate $\eeff$ of the anomalous events.

\begin{figure}[hbpt]
    \centering
    \includegraphics[width=0.45\textwidth]{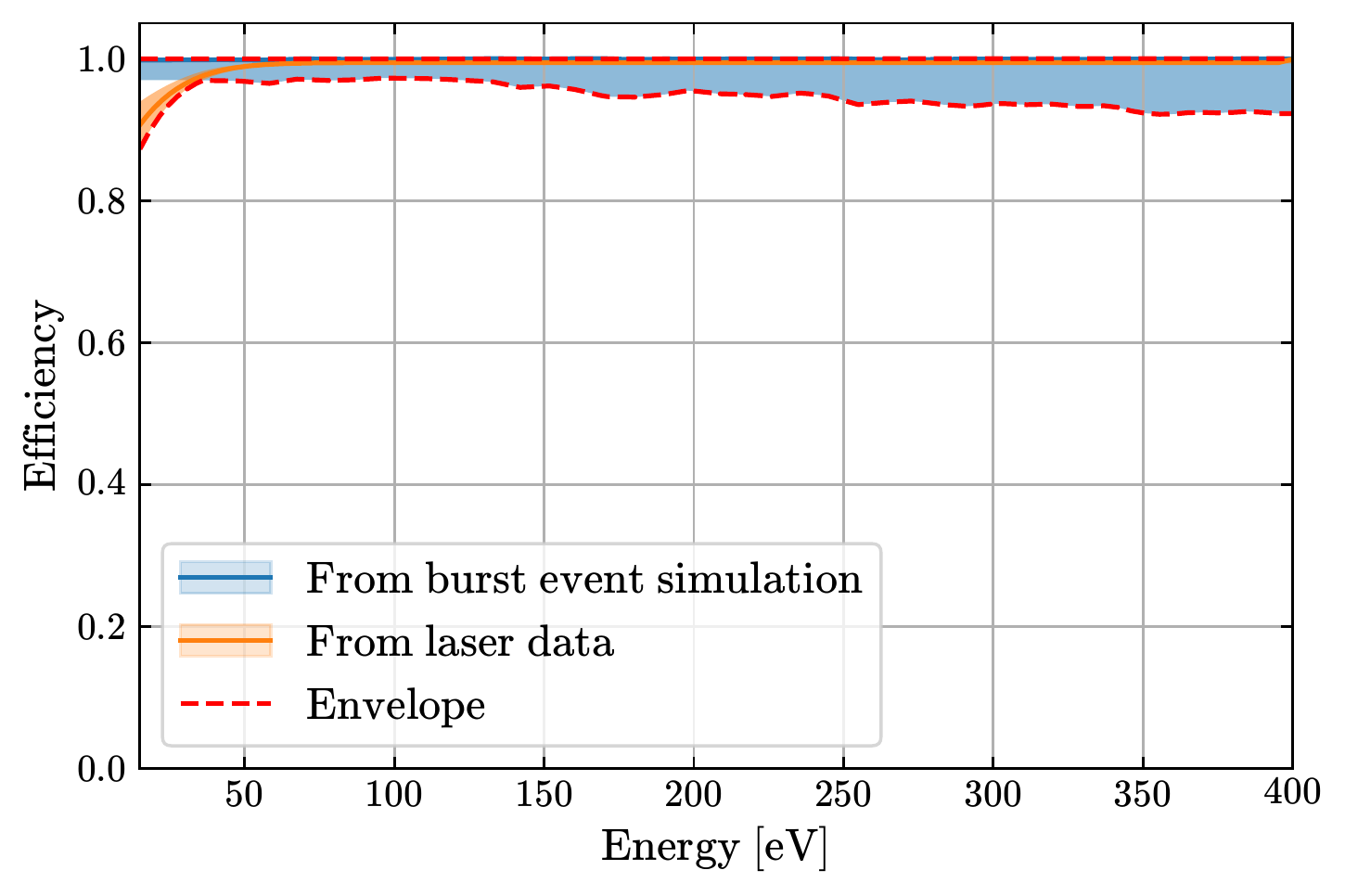}
    \caption{0 V data selection efficiencies evaluated with laser calibration data (orange) and burst event simulation (blue), and associated uncertainties (shading). The dashed red lines are the envelope of the two uncertainty bands, which is used as the total uncertainty of the selection efficiency.}
    \label{fig:cuteff}
\end{figure}

Before comparing the energy spectra, we correct the energy spectra for their event-selection efficiency. We evaluated the selection efficiency of the $\chi^2$ and pulse width selections in the region of $25-150$\,eV of reconstructed recoil energy. We expect the 0V data to be a mix of both calibration-like events and the long-tailed events. The selection efficiency is thus evaluated on both the laser-calibration data and burst event simulation. We estimate the uncertainty for the latter from the three simulated secondary-pulse scenarios. We estimate the selection efficiency as the combination of the two efficiency curves and assign their total uncertainty as the systematic uncertainty (see Fig.~\ref{fig:cuteff}).
We note that for the corresponding energy region in the 60\,V and 100\,V data, the selection efficiency evaluated with the burst event simulation is 100\%.

\begin{figure}[htbp]
    \centering
    \includegraphics[width=0.45\textwidth]{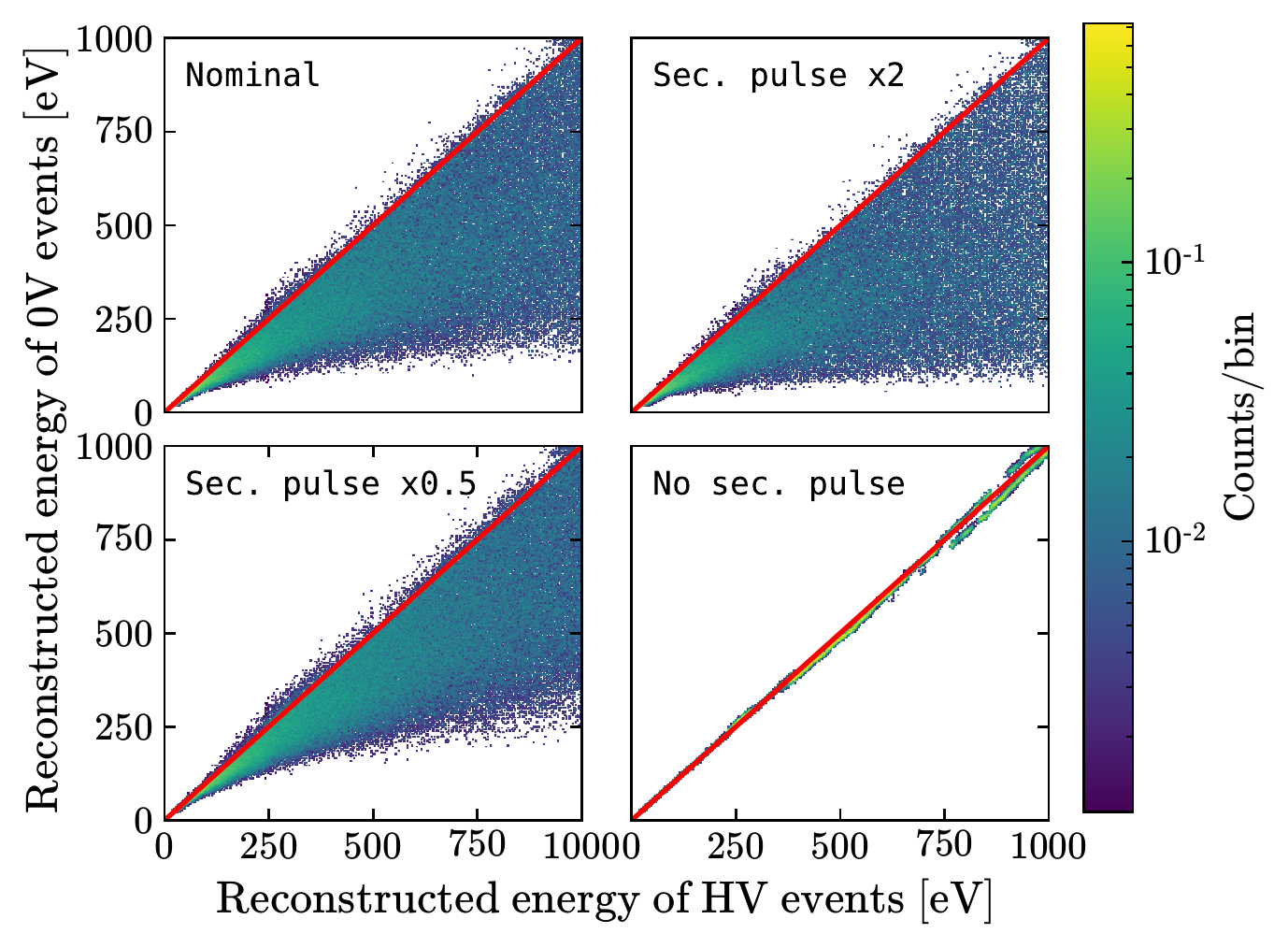}
    \caption{Response matrices that convert the 100\,V spectrum to 0\,V assuming $\eeff=4\,$eV. The four panels correspond to different settings for the rate of secondary pulses in the burst event simulation.}
    \label{fig:respose_matrix}
\end{figure}

We then use response matrices to correct for the detector response difference between the HV mode and the 0V mode.
The response matrices quantify the probability density function of an event being reconstructed in an energy bin with high voltage applied, provided that it is observed in a specific energy bin in the 0V data. The response matrices are evaluated with the burst event simulation. 
For each event in the simulation, traces at 0\,V, 60\,V, and 100\,V are generated with $\eeff$ from 2-7\,eV in steps of 0.5\,eV. We processed the events at different voltages with the same algorithms as the detector data, and use the 2D histogram of the reconstructed energy of HV events versus 0V events to build response matrices.
Examples of response matrices with the three different $N_\mathrm{s}$ models as described in Sec.~\ref{sec:6-compare}~B are shown in Fig.~\ref{fig:respose_matrix}, which also shows a fourth response matrix estimated from a simulation sample with no secondary pulses.
We perform the correction by multiplying these matrices with the uncorrected recoil energy spectra. For each HV-mode spectrum, we assign an envelope corresponding to the spread of the spectra calculated with the four matrices as the systematic uncertainty for the correction.

\begin{figure}[htbp]
    \centering
    \includegraphics[width=0.45\textwidth]{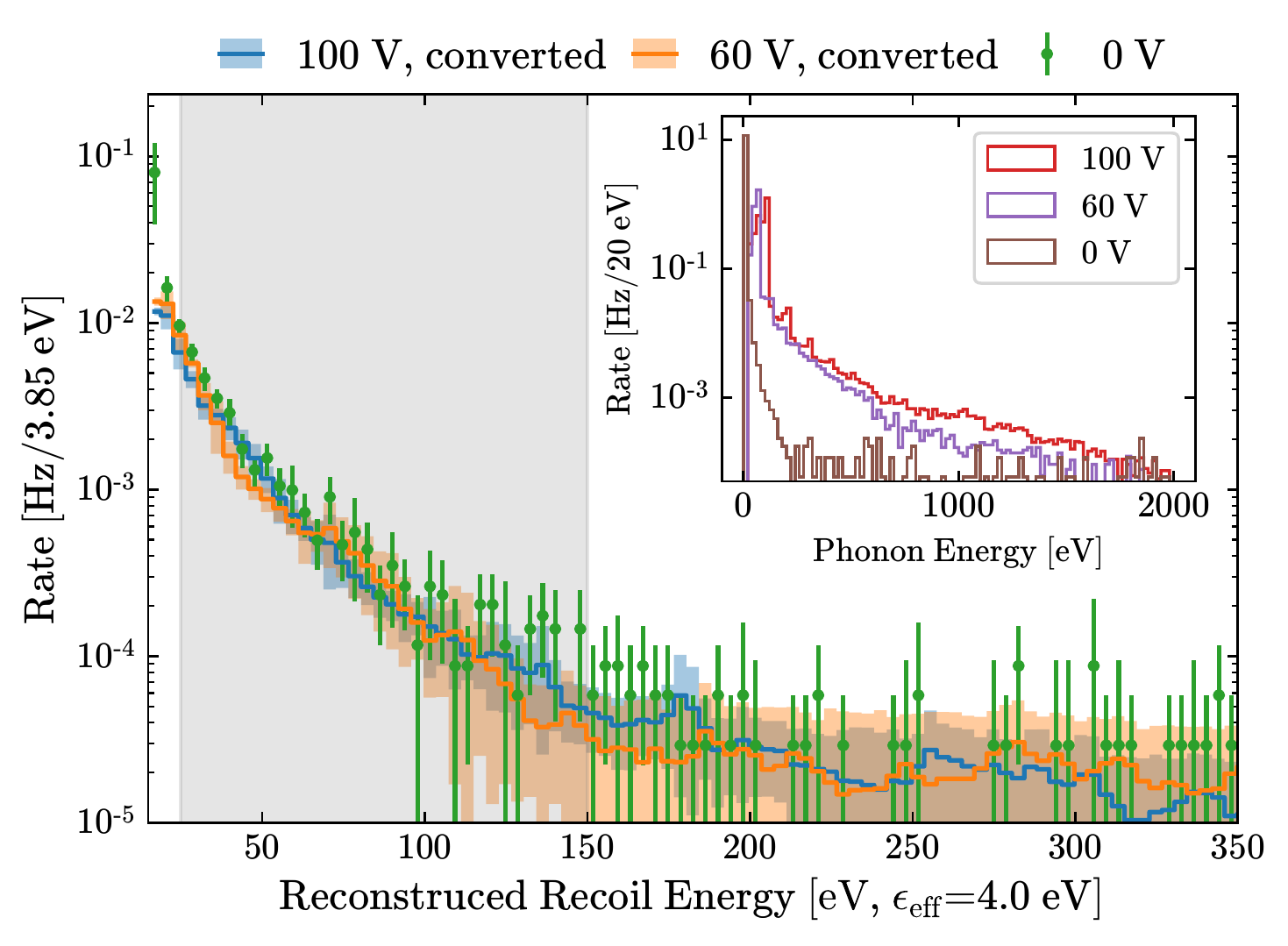}
    \caption{Comparison of the converted HV spectra with the 0V spectrum. The gray area shows the energy range (25-150\,$\mathrm{eV}$) where the $\chi^2$ is calculated. The inset plot shows the phonon spectra before applying the response matrix conversion.}
    \label{fig:HV-0V-spectra}
\end{figure}

Finally, we scan over $\eeff$ and compare the goodness of the fit ($\chi^2$) between the converted HV spectra and the 0V spectrum in the recoil energy region of 25\,eV to 150\,eV. Figure~\ref{fig:HV-0V-spectra} shows an example of the 0V spectrum along with the converted 100\,V and 60\,V spectra at $\eeff=4\,\mathrm{eV}$. 
We find that the converted HV spectra best match the 0V spectrum for an $\eeff$ of 4-5\,eV, with a shallow minimum in $\chi^2$ for these averaged \eh production energies. We note that the $\chi^2$ does not take into account the correlation and systematic uncertainties, thus we are not reporting the exact minimum and uncertainties of $\eeff$. Figure~\ref{fig:HV-0V-spectra} also shows the spectra before the conversion in the inset panel.

\section{Discussion}\label{sec:discussion}

The comparison of the pulse shapes and energy spectra in Sec.~\ref{sec:6-compare} suggests that the HV and 0V background may be dominated by events from the same origin. In this section, we discuss a model that is consistent with these observations drawing from the information in Sec.~\ref{sec:6-compare} and additional circumstantial evidences.

\subsection{A possible explanation of the burst events}

In Sec.~\ref{sec:6-compare}~C we showed that the primary pulse of burst events has an $\eeff$ around 4-5\,eV, with the assumption that the HV and 0V background events have the same predominant origin. There are at least two possible mechanisms that will result in an $\eeff$ close to 4-5\,eV: 1) a single electron recoil event with an energy higher than 20\,eV, which will have $\eeff=3.8\,\mathrm{eV}$; 2) a group of sub-10\,eV electron recoil events that all occur within a couple of \SI{}{\micro\second} time scale (and thus look like a single higher-energy pulse) can have an $\eeff$ around 4-5\,eV according to Ref.~\cite{scholze2000determination}.

Furthermore, we found that the luminescence effect can explain what we have observed assuming that the primary pulse is a collection of 4-5\,eV events. 
For example, $\mathrm{SiO_2}$, the primary component of the PCB that holds the detector, can create luminescence photons of 4.4\,eV, 1.9\,eV, and 2.7\,eV with a decay time of \SI{1.5}{\micro\second}, \SI{20}{\micro\second}, and $7\,\mathrm{ms}$, respectively~\cite{trukhin2003cathodoluminescence,baraban2019luminescence}. The energies and time scales of the 4.4\,eV and 2.7\,eV photons are consistent with the results of Sec.~\ref{sec:5-pulse_shape} and Sec.~\ref{sec:6-compare}. The time constant of the 1.9\,eV photons is close to the pulse fall time in our detector, and can be reconstructed as part of the primary pulse.

Besides luminescence, Cherenkov radiation and transition radiation have been suggested as possible sources of the low-energy excess seen in DM searches with an ER signal~\cite{du2020sources}. We do not evaluate these two mechanisms here because they will not produce a chain of events on the time scale of \SI{}{\micro\second} observed by our dominant source of background events, the burst events. They may become important once we can eliminate burst events.

\subsection{Slow events}

\begin{figure}
    \centering
    \includegraphics[width=0.45\textwidth]{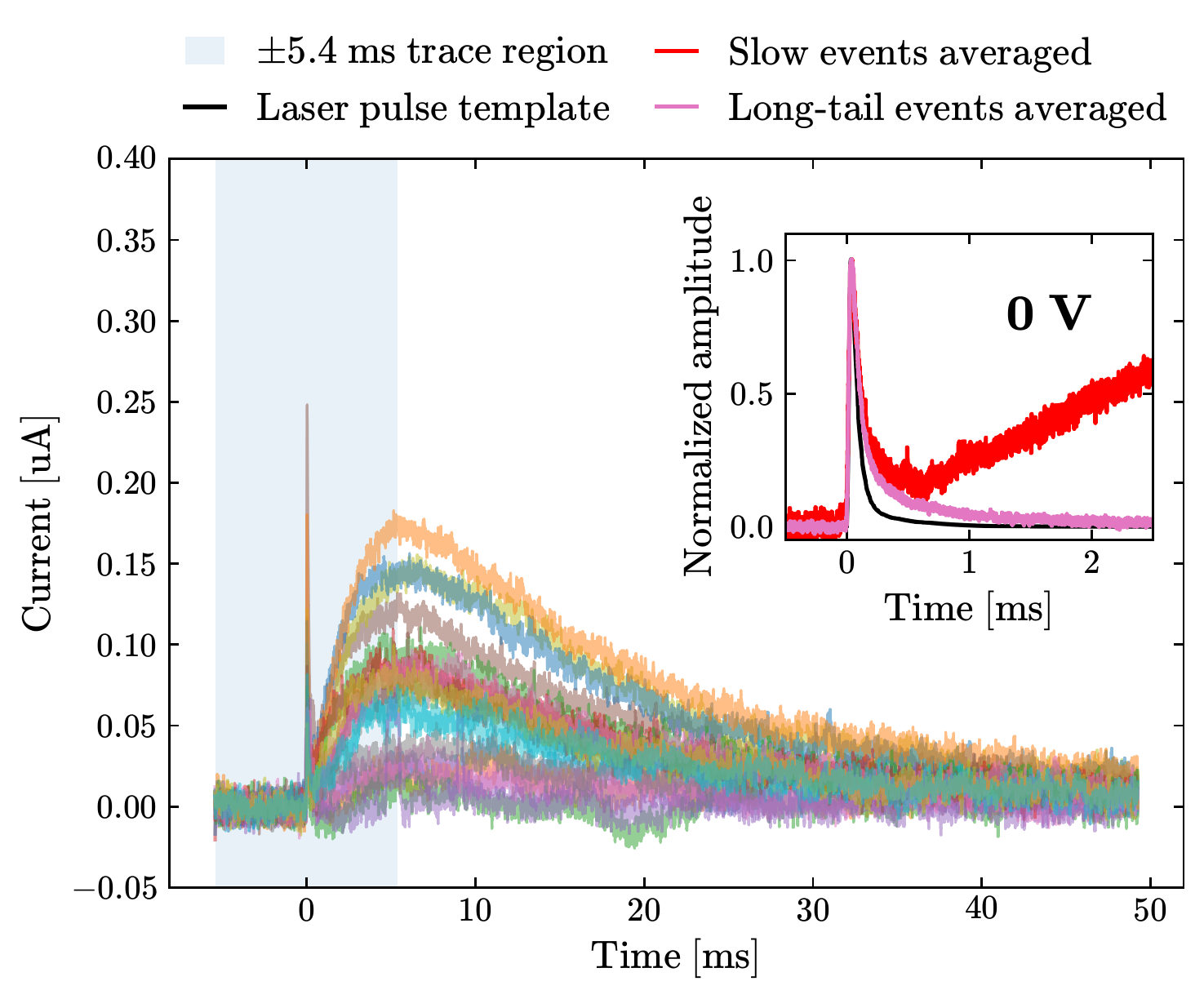}
    \includegraphics[width=0.45\textwidth]{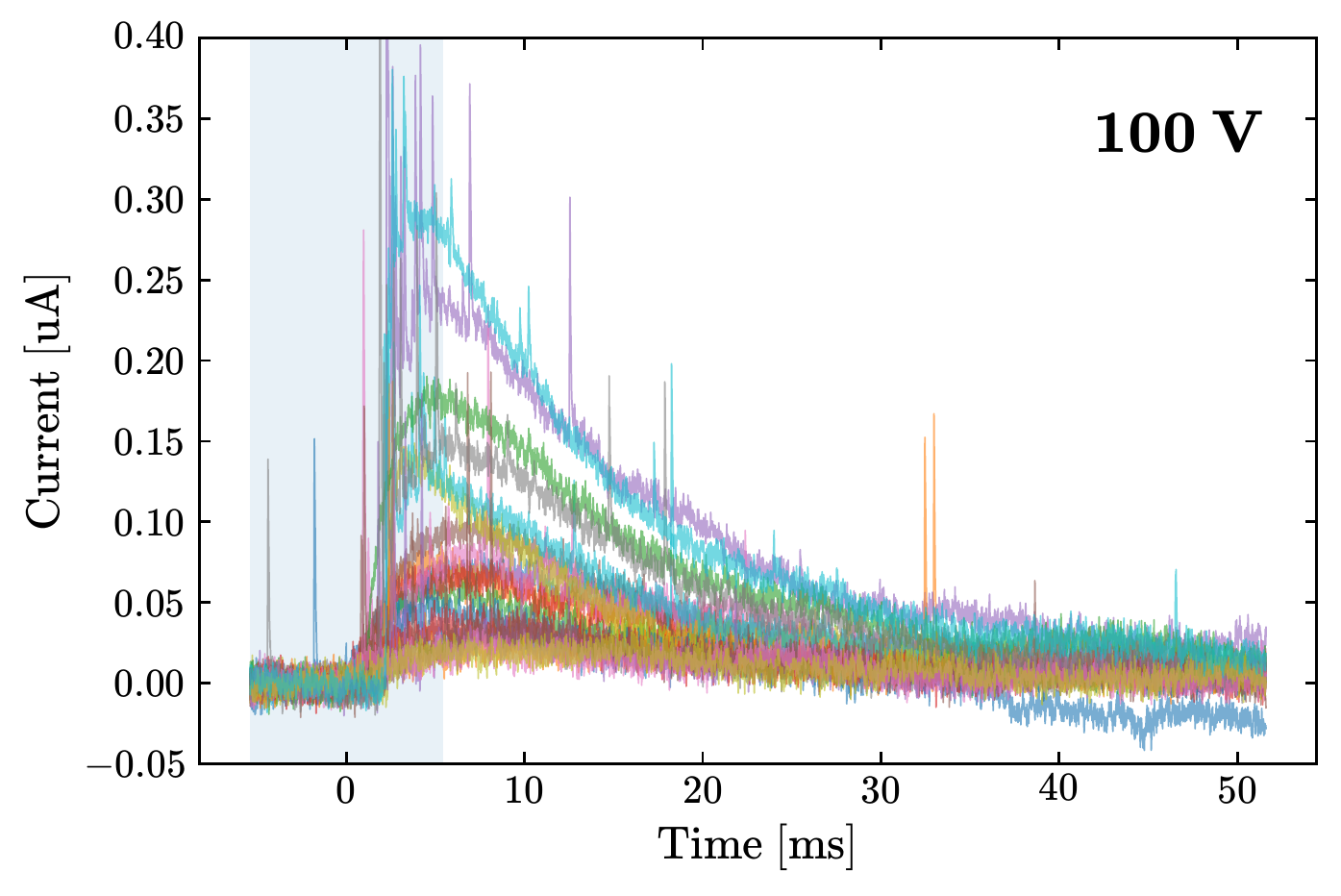}    
    \caption{Example ``slow events" that exhibit a second, slow pulse from 0\,V (top) and 100\,V (bottom) data. The shaded region shows the standard trace length that has been used elsewhere in this paper. The slow pulses extend far beyond the regular trace length. The inset plot of the top panel shows the zoom-in of the averaged pulse shape of the 0\,V slow events in the main plot, compared with the averaged 0\,V long-tail events (pink) as in Fig.~\ref{fig:longtail_selection} and the laser-pulse template.}
    \label{fig:slow_pulse}
\end{figure}

Interestingly, we also noticed a group of events in the 0\,V dataset with a large slope in the pulse during the 5.4\,ms post-trigger region. Upon further investigation, we found that all of these events have a long-timescale pulse with fall time $>10$ ms following the initial pulse. Similar events also appear in 100\,V data, as shown in Fig.~\ref{fig:slow_pulse}. We refer to these events as ``slow events".
We note that the first, fast pulses of the 0\,V slow events have an average shape compatible with 0\,V long-tail events within 0.5~ms, as shown in the inset plot in Fig.~\ref{fig:slow_pulse} top panel.
We also note that about one third of the 100\,V slow events are accompanied by a series of single \eh size pulses, while the slow pulses are of similar sizes like those in the 0\,V data.

The slow pulses could be from energy deposition of high energy particles in the detector holder PCBs of which we would expect a much longer time constant than of energy depositions in the detector directly.  The energy deposition in the PCB may generate luminescence photons, some of which might then be absorbed in the HVeV detectors, causing slow pulses with single \eh burst events as seen in the HV data. In 0V data these would show up as long-tail events combined with the slow pulses, consistent with our observation. The presence of these slow pulses with single \eh burst events is then consistent with the luminescence explanation of the burst and long-tail events.

\section{Conclusion}
In this paper we have presented an analysis of data taken with a SuperCDMS-HVeV detector operated at 0\,V, 60\,V and 100\,V. We obtained a dark matter limit with the 0V exposure, which benefited from the low energy threshold of this detector. The dark matter limit is competitive even with the very small exposure of 0.19\,gram·days. We investigated the low-energy events in the dark matter search data at all three bias voltages. We have shown that both our 0V and HV data can be explained by a common scintillation-like source of background events that have an \eh creation energy of 4-5 eV and are followed by time-correlated bursts of secondary excitations.
We consider luminescence from the detector holder material to be a likely origin of these excess events. We have designed a new detector holder which minimizes insulator material inside the detector volume to reduce this potential background in our next science campaign.
However, with the existing data we cannot rule out the alternate hypothesis that the HV burst events are induced by the voltage, and the 0V long-tail events are caused by a different unknown source.

\begin{acknowledgements}
We thank Noemie Bastidon for her work in
the preliminary design of our optical fiber setup and
wire bonding. We gratefully acknowledge support from
the U.S. Department of Energy (DOE) Office of High
Energy Physics and from the National Science Foundation (NSF). This work was supported in part under
NSF Grants No.~1809730 and No.~1707704, as well as by
the Arthur B. McDonald Canadian Astroparticle Physics
Research Institute, NSERC Canada, the Canada Excellence Research Chair Fund, Deutsche Forschungsgemeinschaft (DFG, German Research Foundation) under Project No.\,420484612 and under Germany’s Excellence
Strategy - EXC 2121 “Quantum Universe” - 390833306,
and the Department of Atomic Energy Government of
India (DAE) under the project “Research in basic sciences" (Dark matter). Fermilab is operated by Fermi
Research Alliance, LLC, under Contract No.\,DE-AC02-
37407CH11359 with the US Department of Energy. Pacific Northwest National Laboratory (PNNL) is operated
by Battelle Memorial Institute for the DOE under Contract No.\,DE-AC05-76RL01830. SLAC is operated under
Contract No. DEAC02-76SF00515 with the United States
Department of Energy.
\end{acknowledgements}

\bibliographystyle{apsrev4-1}
\bibliography{refs}

\end{document}